\documentclass[manuscript,screen]{acmart}
\AtBeginDocument{%
  \providecommand\BibTeX{{%
    \normalfont B\kern-0.5em{\scshape i\kern-0.25em b}\kern-0.8em\TeX}}}

\usepackage{subfig
	,listings
}

\usepackage{paralist}
\newcommand{\nd}{\vspace{1mm}\noindent}

\usepackage{tcolorbox}
\newcommand{\emt}[1]{\emph{''#1''}}
\usepackage{enumitem}

\def\bf{\textbf}
\def\it{\textit}
\def\fig {Figure~}
\def\tbl {Table~}
\def\sec {Section~}
\def\secs {Sections~}
\usepackage{tikz}        
\usepackage{pgf-pie}
\usetikzlibrary{positioning,shadows}
\newif\ifpienumberinlegend
\pgfkeys{/number in legend/.code=
    \expandafter\let\expandafter\ifpienumberinlegend
    \csname if#1\endcsname
    \ifpienumberinlegend

    \def\beforenumber##1\afternumber{}%
    \fi,
    /number in legend/.default=true
}

\definecolor{codegreen}{rgb}{0,0.6,0}
\definecolor{codegray}{rgb}{0.5,0.5,0.5}
\definecolor{codepurple}{rgb}{0.58,0,0.82}
\definecolor{backcolour}{rgb}{255,255,255}

\lstdefinestyle{mystyle}{
    backgroundcolor=\color{backcolour}, 
    numberstyle=\color{codegray},
    stringstyle=\color{codepurple},
    keywordstyle=\color{magenta},
    numberstyle=\tiny\color{codegray},
    basicstyle=\ttfamily\footnotesize,
    captionpos=b,                    
    numbers=left,                    
    numbersep=5pt,                  
}

\RequirePackage{expl3}    
\ExplSyntaxOn             
\ExplSyntaxOff            

\begin{document}
\title[A Large-Scale Study 
of IoT Security Weaknesses and Vulnerabilities in Crowd-Shared C/C++ IoT Code Examples]{A Large-Scale Study 
of IoT Security Weaknesses and Vulnerabilities in the Wild}

\author{Madhu Selvaraj}
\email{madhumitha.selvaraj@ucalgary.ca}
\affiliation{%
  \institution{DISA Lab, University of Calgary}
  \country{Canada}
}
\author{Gias Uddin}
\email{gias.uddin@ucalgary.ca}
\affiliation{%
  \institution{DISA Lab, University of Calgary}
  \country{Canada}
}


\begin{abstract}
 
Internet of Things (IoT) is defined as the connection between places and
physical objects (i.e., things) over the internet/network via smart computing
devices. IoT is a rapidly emerging paradigm that now encompasses almost every
aspect of our modern life. As these devices differ from traditional computing,
it is important to understand the challenges IoT developers face while
implementing proper security measures in their IoT devices. We observed that IoT 
software developers
share solutions to programming questions as code examples on three Stack Exchange Q\&A sites: 
Stack Overflow (SO), Arduino, and Raspberry Pi. Previous research studies 
found vulnerabilities/weaknesses in C/C++ code examples shared in Stack Overflow. 
However, the studies did not investigate C/C++ code examples related to IoT. 
The studies investigated SO code examples only. In this paper, 
we conduct a large-scale empirical study of all IoT C/C++ code examples shared in the three 
Stack Exchange sites, i.e., SO, Arduino, and Raspberry Pi.  
From the 11,329 obtained code
snippets from the three
sites, we identify 29 distinct CWE (Common Weakness Enumeration) 
types in 609 snippets. These CWE types can be
categorized into 8 general weakness categories, and we observe that evaluation,
memory, and initialization related weaknesses are the most common to be
introduced by users when posting programming solutions. Furthermore, we find
that 39.58\% of the vulnerable code snippets contain instances of CWE types that
can be mapped to real-world occurrences of those CWE types (i.e CVE instances).
The most number vulnerable IoT code examples was found in Arduino, followed by 
SO, and Raspberry Pi. Memory type vulnerabilities are on the rise in the sites. For example, from the 3595  mapped CVE instances, we find that 28.99\% result in Denial of
Service (DoS) errors, which is particularly harmful for network reliant IoT
devices such as smart cars. Our study results can guide various IoT stakeholders to be aware of such vulnerable IoT code examples and to 
inform IoT researchers during their development of tools that can help prevent developers 
the sharing of such vulnerable code examples in the sites. 
\end{abstract}





\maketitle

\section{Introduction}
Internet of Things (IoT) is an interconnected system of physical objects
("things") via the Internet for the purpose of exchanging information and
communication ~\cite{InternetofThings}. The rapid developments in this
technology has made it possible for everyday devices to be connected to the
internet, and is now unavoidable in modern life. At the end of 2020 there were a
total of 11.3 billion connected IoT devices, and rapid expansion is expected to
continue with an estimated 27.1 connected IoT devices by 2025
~\cite{StateofIoT}. The increasing prevalence of IoT devices such as home
security systems, cars, and smart TVs being a part of everyday life has
consequently increased the risk of IoT security vulnerabilities and threats.
Such devices can face unique security challenges such as requiring large dynamic
networks which increases the attack surface area ~\cite{SmartHomesSecurity},
data profiling by third-party developers ~\cite{SmartHomePersonalAssistants},
and man in the middle attacks ~\cite{IoTPrivacySecurity}. As increasing demand
is being placed on developers to release devices more quickly, lower security
standards and poor coding practices are likely to become more frequent
~\cite{SmartHomesSecurity}. Therefore, it is important to understand the
particular risks associated with IoT devices and their development.

Developers of IoT devices, similar to developers of any software product, often
refer to online Q\&A sites such as those part of the Stack Exchange network for
solutions to their problems. The answers posted on these sites often contain
code examples as part of the answer. The Stack Exchange site Stack Overflow (SO) in
particular contains code snippets in 75\% of their answers
~\cite{SOcodesnippets}. These code examples are then often directly reused by
developers without modifications. As found in previous studies, these code
examples can also contain security vulnerabilities. For example, when studying
Python answers posted on Stack Overflow, Rahman et al. observed that 9.8\% of
7,444 Stack Overflow accepted answers contained at least one instance of a poor
coding practice, with code injection occurring the most frequently
~\cite{PythonVulnerabilities}. Fischer et al.~\cite{cFISCHER} found that vulnerable Android code examples shared in 
SO are reused in millions of popular Android apps.

C/C++ is the fourth most popular language in the world are widely used in IoT devices~\cite{cTIOBE}. 
In the CVE (Common Vulnerabilities Exposure) database, 49\% of the vulnerabilities are related to C/C++ programming 
language. In two separate studies by Verdi et al.~\cite{EmpiricalC++Study} and Zhang et al.~\cite{C/C++SO} found 
that C/C++ code examples shared in SO can have critical security vulnerabilities.
For example, Zhang et al. observed 24,803 instances of 32 difference types of code weaknesses
in 11,748 code snippets ~\cite{C/C++SO}. The studies offer valuable insights about the weakness/vulnerabilities in the shared SO C/C++ code examples. 
However, the studies did not focus on the IoT code examples only. While both studies analyzed \textit{all} C/C++ code examples which should also include 
the IoT-related C/C++ code examples, all the results are not directly generalizable to the IoT domain. For example, some vulnerabilities might not exist in IoT code examples, while some others may be comparatively more visible in IoT code examples. 
Zhang et al.~\cite{C/C++SO} found CWE types such as CWE 775 - Missing release of file descriptor or handle after effective lifetime and CWE 910 - Use of expired file descriptor in their analysis. On the other hand, we detected some CWE types that were not found in their analysis, such as CWE 595 - Comparison of object references instead of object contents. An example of this shown in Listing \ref{595-ex}, where instead of comparing content to "download" using the operator "==", strcmp() should have been used to do a proper comparison of object contents. 
\begin{figure}[t]
\begin{lstlisting}[language=C++,label=595-ex,caption=CWE 595 - Comparison of object references instead of object contents]
String content;
void loop()
{  
    while(Serial1.available()) {
       character = Serial1.read();
       content.concat(character);
     }

if (content == "download")
{
    Serial1.println(content);
    content = "";
    read_from_SD_Card();
    }

........................
    ........................
    ........................
}//end of loop()
\end{lstlisting}
\end{figure}
Second, a qualitative analysis of the vulnerable code examples 
is absent in the previous papers (e.g., categorization of the vulnerabilities, examples of vulnerable code examples, etc.) ~\cite{EmpiricalC++Study} ~\cite{C/C++SO} .
Third, both papers only studied SO, while we observed that IoT code examples are shared across multiple Stack Exchanges sites. 
For example, Stack Exchanges sites like Arduino, Raspberry Pi, IoT, and IoTa focus exclusively on IoT-based Q\&A.
    



In
this paper, we study all the IoT C/C++ code snippets shared in
three different Stack Exchange sites: SO, Arduino, and Raspberry Pi. 
We found total 11,329 IoT code snippets in the three sites. By analyzing the code snippets for security weakness, we answer four research questions (RQ). 

\nd\textbf{RQ1. What are the 
different types of weaknesses found in the shared IoT code examples?}\newline 
Vulnerabilities can be introduced into source code, if the code reused from online forums can exhibit weaknesses.
The CWE (Common Weakness Enumeration) database contains a list of poor/weak coding pattern that 
can introduce vulnerabilites into source code. We apply static parser Cppcheck \cite{cppcheck} to each code snippet, 
which returns whether the snippet has any weakness that matches with the CWE database of patterns. 
Cppcheck also returns the CWE ID to denote the weakness. By applying Cppcheck on all 11,329 code examples, 
we found 609 \it{weak} code snippets. The 609 weak code snippets are found in three of the five studied sites: Arduino, SO, and Raspberry Pi.
Arduino contained the most number of non CWE-398 weaknesses (123), followed by SO (97). 
The weak snippets corresponded to 29 distinct CWE types. The most prevalent weakness is CWE 398, i.e., poor coding quality (in 422 snippets). 
A code snippet may exhibit more than one weakness types. 
Total 240 weak code snippets exhibited non CWE-398 weakness that belong to 28 CWE types. 
We group the 28 CWE types into eight ``weakness'' categories by analyzing the description of the types: 
Function, Memory, Evaluation, Conversion, Resource, Calculation, Reachability, and Initialization.
Function type weaknesses are found with the most number of CWE IDs (7), followed by Memory type weaknesses (6). 
%
%

\nd\textbf{RQ2. How do the weakness types map to
common vulnerability exposure (CVE) instances?}\newline 
CVEs (Common Vulnerabilites Exposures) are instances of
CWE types that have occurred in real-world software. A given CWE type can map to zero, one or more than one CVE. 
We map each of the 28 distinct CWE types to their CVE instances using
cvedetails.com. We find that 12 of the CWEs map to one or more CVEs. Total 95 code snippets belong to these 12 CWE types
and they are mapped to a total of 3595 CVE instances. Out of the 12 mapped CWE types, the most (4)  
belong to the Memory category as defined in RQ1. These 4 CWEs map to total 2997 CVEs. 

\nd\textbf{RQ3. How are the mapped vulnerabilities
classified in the CVE details database?}\newline cvedetails.com groups 
the CVEs into 13 different types. We
analyze which specific CVE types as mapped to our 12 CWEs from RQ2. We find that the CWEs belong to eight 
CVE types in cvedetails.com. The most common type of vulnerability is Overflow (1308 CVEs), followed
by Denial of Service (1213 CVEs) and Code Execution (464 CVEs). 

\nd\textbf{RQ4. How do the weakness types evolve over time in the shared IoT code examples?}\newline
Analyzing the evolution of the vulnerabilities detected in the Stack Exchange
code snippets will provide us with a better understanding of concerning trends
regarding particular types of weaknesses. To observe the evolution of the 8
general weakness categories from RQ1, we analyze the number of weak code
snippets posted each year. We find that in general the number of weak code snippets related to evaluation, initialization, and memory errors has increased.

Our study results show that most of the shared IoT code examples do not have any known security weakness/vulnerability and only 
2.1\% of all shared IoT code examples have security weaknesses. While we studied five sites from Stack Exchange, we found weak code snippets 
in three of those (Arduino, SO, and Raspberry Pi), with Arduino containing the most number of weak code snippets. This is not surprising, given 
sites like Arduino are dedicated to IoT-based Q\&A only. Our study results can be used by security practitioners to determine the 
severity of reusing IoT C/C++ code examples from online forums. Security researchers can use our findings to prioritize their efforts 
on the types of tools and techniques that can be developed to warn IoT developers to assist them during sharing. This is important given 
no single tool can be successful to analyze all the different weakness types.

\textbf{Replication Package: }\url{https://github.com/disa-lab/IoTCodeWeaknessStackExchange}

\section{Background}
Our study data are collected from Stack Exchange sites. We offer a brief background info on the Stack Exchange network sites in \sec\ref{sec:background-sesites}. 
Our study relies on CWE (Common Weakness Enumeration) and CVE (Common Vulnerability Exposure) databases. We briefly explain the fundamental concepts behind CWE and CVE databases in \secs\ref{sec:background-cwe}, \ref{sec:background-cve}.
\subsection{Stack Exchange Sites}\label{sec:background-sesites}
 
The Stack Exchange network of sites is a series of Q/A websites that covers
topics from a variety of different fields. This network consists of 172
different sites that were accessed by more than 100 million users
~\cite{SOSites} ~\cite{SOUsers}. Sites that are related to programming and
technology are often a common resource for developers, and answers containing
code snippets are frequently reused and copied directly into real-world software
such as Android applications ~\cite{Abdalkareem2017OnCR} 
~\cite{SOCodeLaundering}. Questions are posted on these sites by users of
varying skill levels, are then answered by users of the same community.
Depending on the nature of the question, answers can contain code examples along
with the text response. As Stack Exchange sites attract many types users,
including professional developers and beginners, the quality of answers can vary
greatly. Higher quality answers are highlighted on these sites by either the
user that asked the question selecting a particular answer to be accepted, or by
any user up-voting a answer ~\cite{ReadingAnswersSO}. These up-votes contribute
to a user's reputation score, and Rahman et al. observed that this score does
not correlate with answer quality, as both high and low rated users are equally
likely to introduce vulnerabilities ~\cite{PythonVulnerabilities}.

\subsection{Common Weakness Enumeration (CWE)}\label{sec:background-cwe}
\begin{figure}[t]
\centering
\includegraphics[scale=0.22]{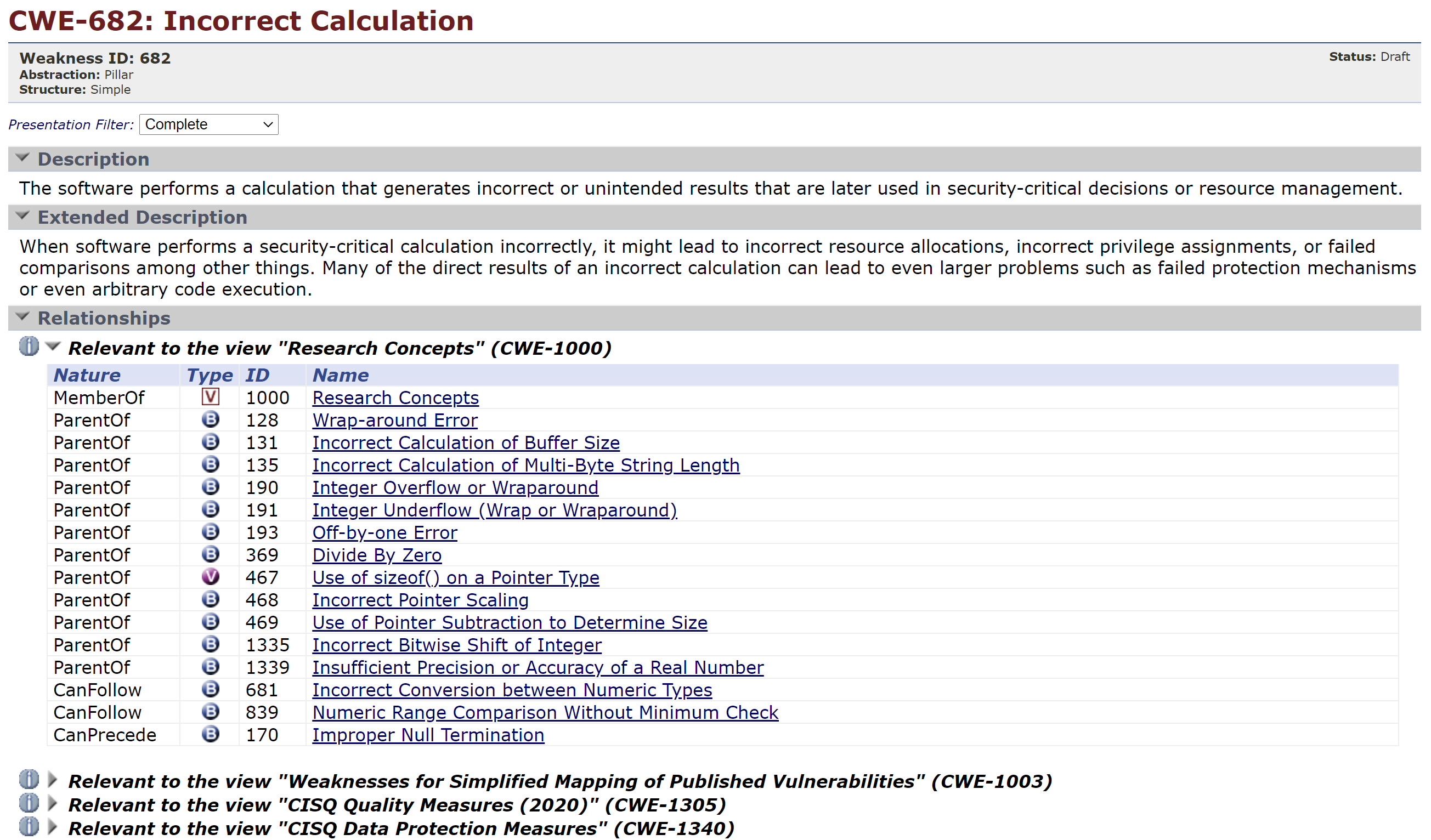}
\caption{Screenshot of an entry in the CWE database}
\label{fig:cwe_database}
\end{figure}
\begin{figure}[t]
\centering
\includegraphics[scale=0.40]{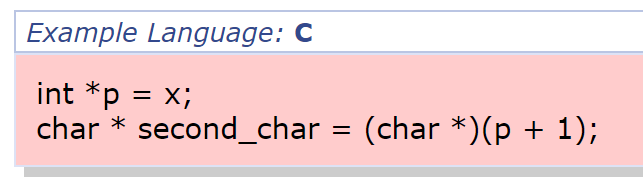}
\caption{Example of a CWE instance}
\label{fig:cwe_database-instance}
\end{figure}

Common Weakness Enumeration (CWE) is a community based list of software
weaknesses maintained by the Mitre Corporation. It's goal is to make
identification of vulnerabilities in software easier, and to prevent errors in
products before release ~\cite{CWEDef}. The list is consistently updated with
new weaknesses. As of July 2021, there are a total of 924 weaknesses in CWE
Version 4.6 with 92 related to C/C++ ~\cite{CWEVersion} ~\cite{C_weaknesses}
~\cite{C++_weaknesses}. Figure ~\ref{fig:cwe_database} shows an example of an
entry in the CWE list, also known as a CWE type. Each entry contains further
information about the CWE type, and we see that for CWE 682 - Incorrect
Calculation, some important details include a detailed description with common
consequences of the vulnerability, and it's relationships to other CWE types. We
observe that CWE types can have a parent-child relationship with each other, and
in this case CWE 682 - Incorrect Calculation is a parent of CWE types that
represent specific calculation errors, such as CWE 369 - Divide by Zero. 

An entry in the CWE database also contains examples with explanations as to why
they harmful. We see in the example shown in Figure
~\ref{fig:cwe_database-instance} that instead of calculating the address of the
second byte of p to assign to second\_char, "p+1" instead results in the
addition of sizeof(int) to p. An incorrect calculation such as this example can
lead detrimental consequences as memory will be accessed unintentionally,
leading to a potential information leak. 
\subsection{Common Vulnerabilities and Exposure (CVE)}\label{sec:background-cve}
\begin{figure}[t]
\centering
\includegraphics[scale=0.3]{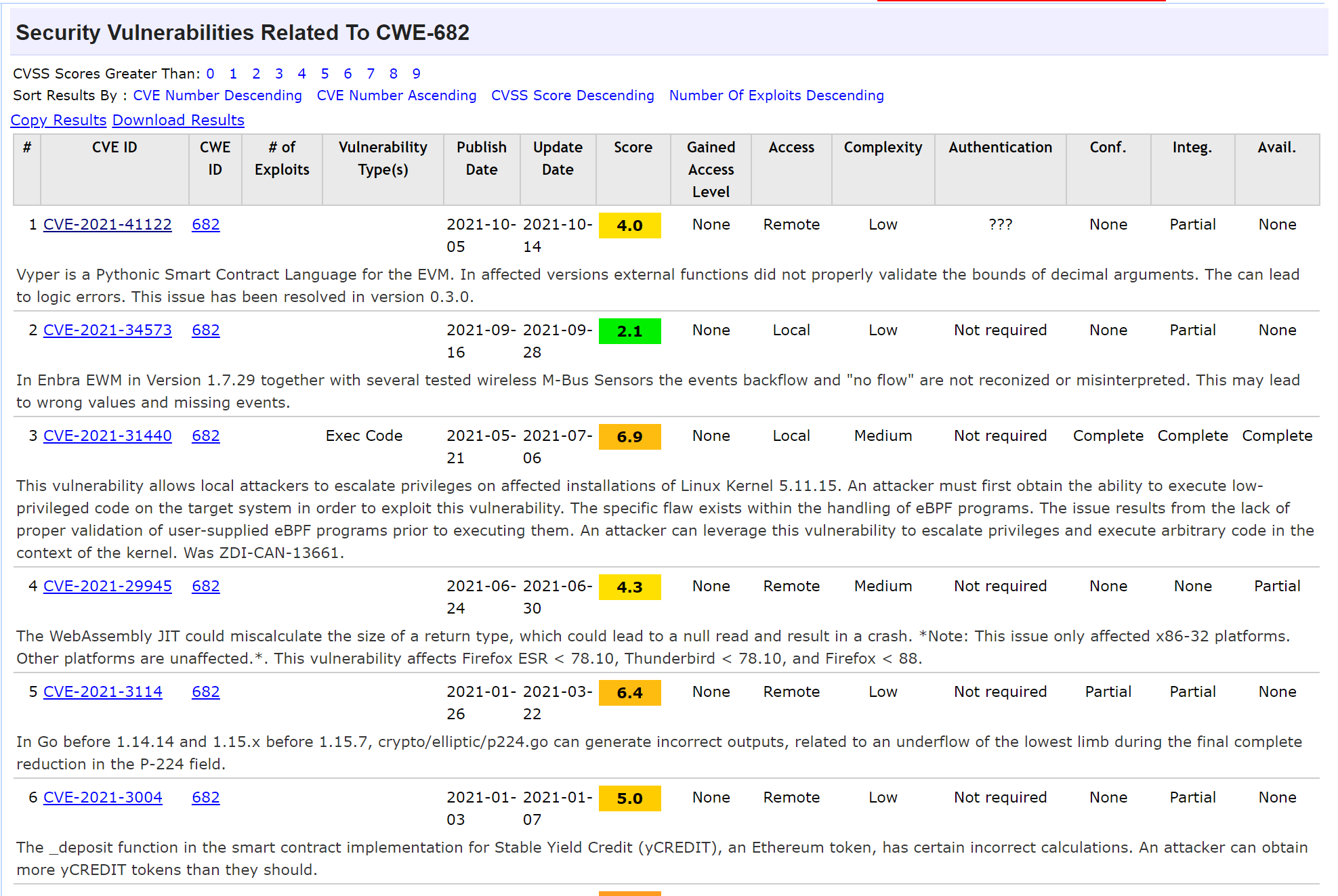}
\caption{Screenshot of CVE instances for CWE 682 - Incorrect Calculation as found on cvedetails.com}
\label{fig:cve_database}
\end{figure}
\begin{figure}[t]
\centering
\includegraphics[scale=0.3]{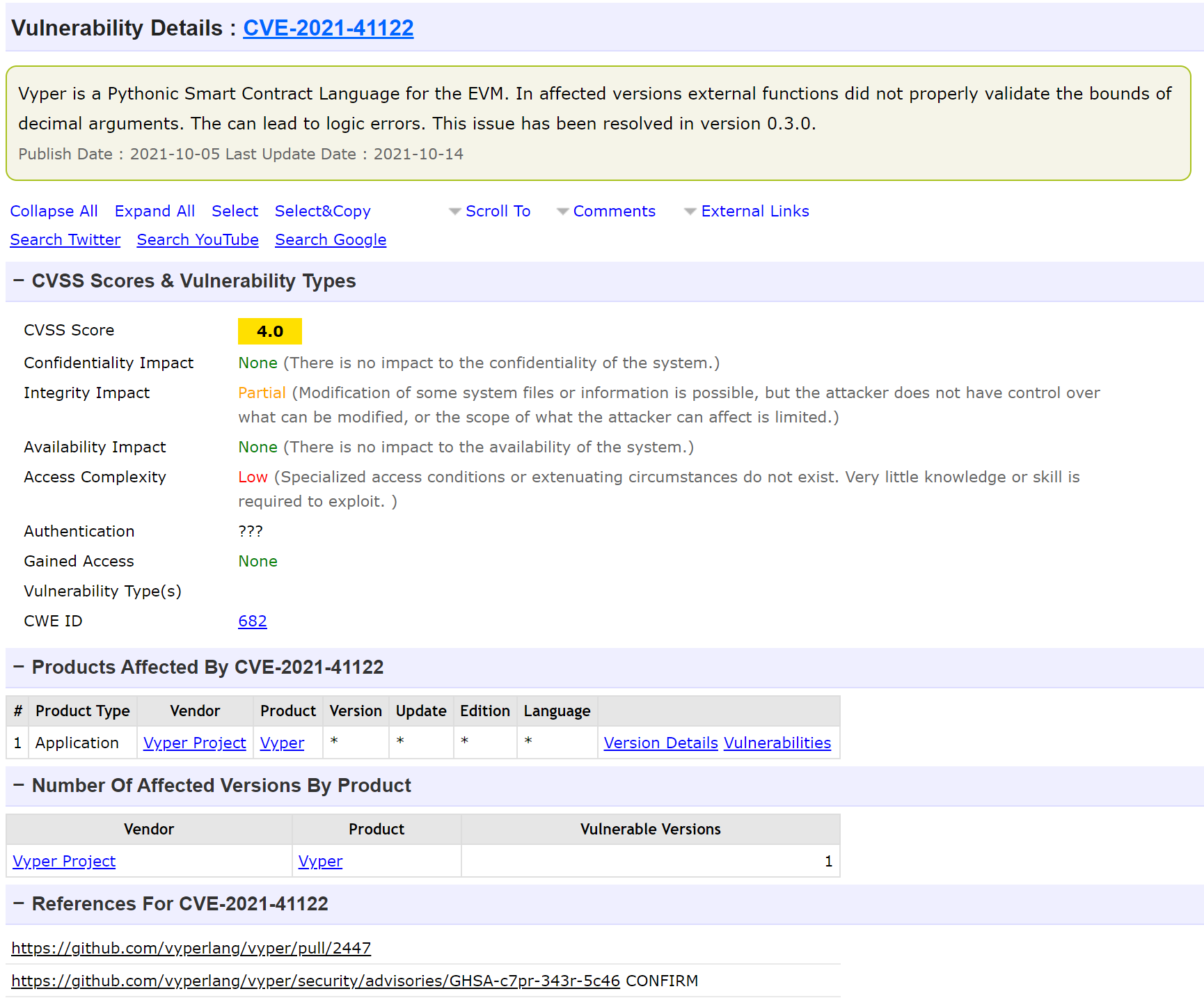}
\caption{Example CVE instance for CWE 682 - Incorrect Calculation }
\label{fig:cve_database-instance}
\end{figure}
Common Vulnerabilities and Exposures (CVE) is a database of instances of CWE
vulnerabilities that have occurred in real-world software applications. It
represents publicly disclosed cybersecurity incidents. This database is maintained by the Mitre Corporation but it
also listed as part of the National Vulnerability Database (NVD). In this paper,
we obtain CVE information from cvedetails.com, which uses the CVE list from the
NVD and provides further insightful information about a CVE instance. The most
important aspect of a CVE instance is it's mapping to a specific CWE type, as
this provides more detail on the real-world implications of software weaknesses.
The mapping between a CVE instance and a CWE type is primarily done by keyword
matching, and the majority of CVE instances are mapped by the National Institute
of Standards and Technology (NIST) ~\cite{mapping-guide}. 

Figure
~\ref{fig:cve_database} shows the first 6 CVE instances that were mapped to CWE
682 - Incorrect Calculation, which has a total of 43. We see that along with the
mapped CWE type, further information is provided in the entry including a
description, a CVSS score, and a vulnerability type. The vulnerability type is
determined by cvedetails.com using keywords from the description, while the CVSS
scoring system is used to assess the severity of a particular instance
~\cite{cvss}. We see in Figure ~\ref{fig:cve_database-instance}, each CVE
instance, in this case CVE-2021-41122, can also be viewed and contains more
detailed information including references to it's origination. 

\section{Data Collection and Preprocessing}
We discuss our data collection process for our study in \sec\ref{sec:data-collection}. 
We then discuss how we preprocess the study data to find security weaknesses in the IoT code examples found in our dataset (\sec\ref{sec:data-preprocessing}).
\subsection{Data Collection}\label{sec:data-collection}

While SO is the most popular Q\&A site for software developers of all kinds (including IoT programmers) in the Stack Exchange network of sites, 
there are four other sites that are specifically setup to foster IoT-based discussions: Arduino, Raspberry Pi, IoT, IoTa. 

First, we downloaded the entire data dump of each of the five sites from the Stack Exchange official dump as of Sept 2021 \cite{StackDump}. Each download consists of multiple xml files consisting of various information about each site and it's contents. The specific files we make us of are Posts.xml and PostHistory.xml. 
Second, we move on to collecting code examples from each site's Posts.xml file. 
In each of the studied stack exchange sites, users are allowed to post and
answer questions. Responses to questions can consist of both plain text and code
examples. The code examples across all five studied sites were consistently
identified by the "pre" HTML tag, which was used to specifically parse out the
code segments for further analysis. Because this study is focused on studying
C/C++ code examples related to IoT, the code snippets across all six sites were
subject to two general filtering criteria: is it C/C++ code and is it related to
IoT. However, the specific filtering process varied according to the individual
site's characteristics. The specific code collection process is broken down below, and is divided into the process followed for SO posts, and for posts from the IoT-specific sites.
 \newline
\subsubsection{Stack Overflow}
To determine if a SO code example was written in C/C++, 
the question tags were checked to see if they contained the
keywords C or C++. Then, since SO is more general and contains a large number of posts un-related to IoT, we filter for IoT related posts by checking if tags contain at least one of 75 IoT-related tags. The list of 75 IoT tags is taken from Uddin et al  ~\cite{Uddin-IoTTopic-EMSE2021}, and a sample is shown in Table \ref{tableIoTKeywords}.
\begin{table}
    \centering
     \caption{List of some IoT related tags used to filter for IoT related Stack Overflow posts}
    \label{tableIoTKeywords}
    \begin{tabular}{llll}
    \hline
    arduino-due & iot & audiotoolbox & audiotrack\\
    arduino-mkr1000 & arduino-uno-wifi & aws-iot & aws-iot-analytics\\
    gpio & azure-iot-edge & azure-iot-hub-device-management & lora \\
    adafruit & android-things & attiny & avrdude \\\hline
    \end{tabular}
\end{table}
\subsubsection{IoT-specific sites (Arduino, Raspberry Pi, IoT, IoTa)}
For the four IoT-specific sites, we observed that in many questions, the tags did not contain programming language related keywords. Therefore, in order to filter for C/C++ code examples we used the language detection tool Guesslang which has an accuracy of 90\% according to its
documentation \cite{guesslangdoc}. We then did not apply any IoT filtering and continued the collection process with all posts as we assumed all were IoT related. 
\newline \newline
Finally to reduce the amount of snippets containing pseudocode, we followed an
approach used by previous studies that found the median line count of SO code blocks to be five ~\cite{SOTorrent}, and only selected snippets containing at least five lines ~\cite{C/C++SO}
~\cite{SpottingCodeExamples}. This technique was used for code snippet
collection for all Stack Exchange sites. 

Overall, we collected a
total of 11,329 IoT related C/C++ code snippets. The breakdown for each
individual site can be found in Table ~\ref{fig:studied_sites_data_collection}.
 \begin{table}[h]
        \centering
        \caption{Statistics of each studied site}
        \begin{tabular}{lrrr}\toprule
        
        \bf{Site Name} & \bf{Total Code Snippets in Posts.xml} & \bf{Total Snippets Collected}\\
        \midrule
        Stack Overflow (SO) & 870,674 & 3,599\\
        Arduino & 53,912 & 6,616\\
        Raspberry Pi & 77,205 & 1,114 \\
        IoT & 4,533 & 30\\
        IoTa & 2,145 & 17\\
        \midrule
        Total & 1,003,469 & 11,329\\ 
        \bottomrule
        \end{tabular}
        
\label{fig:studied_sites_data_collection}
    \end{table}
\subsection{Data Preprocessing}\label{sec:data-preprocessing}
The collected snippets were then analyzed for weakness on the CWE 
(Common Weakness Enumeration) List, which is a list of common software security weaknesses. To analyze all snippets we used a 
static code analyzer called Cppcheck, which supports various types of code checks and allows for specific weaknesses to be suppressed.
Cppcheck is a static code analyzer that can detect weaknesses in C and C++ code.
It supports uncommon syntax common in embedded code. Using version 2.4.1
released in March 2021, we were able to detect weaknesses varying from
uninitialized variables to memory leaks. According to Zhang et al, cppcheck is
able to identify 59 out of the 90 code weaknesses that are related to C and C++
~\cite{C/C++SO}. Cppcheck has been used in previous studies and has shown to be
very precise. In the comparison of code analyzers done by Arusoaie et al.,
cppcheck only had a 0.78 false positive rate against the entire test suite of
650 common C/C++ bugs ~\cite{ArusoaieC/C++Analysis}. Furthermore, in a manual
analysis of cppcheck's accuracy done by Zhang et al, 85 out of 100 CWE instances
detected by cppcheck were labelled as accurate with a strong agreement among the
three authors conducting the analysis (Cohen's Kappa of 0.68) ~\cite{C/C++SO}.

\begin{figure}[t]
\begin{lstlisting}[language=C++,label=563-ex,caption=CWE 563 - Assignment to Variable without Use in a Stack Overflow code snippet ]
   void Server::on_readyRead()
    {
        // "while" loop would block until at least one whole line arrived
        // I would use "if" instead
        if(socket->canReadLine()) 
        {
            QByteArray ba = socket->readLine();

            QByteArray response;

            // some code which parses arrived message
            // and prepares response

            socket->write(response);
        }
        //else just wait for more data
    }
\end{lstlisting}
\end{figure}
\begin{table}[t]
        \centering
        \caption{Errors supressed in cppcheck}
        \begin{tabular}{l|l}\toprule
        
        \bf{Criteria Name} & \bf{Criteria Description}\\
        \midrule
        Syntax Error & Errors in the syntax of the code\\
        Unread Variable & Variable is assigned a value but never used\\
        Unused variable or unused stuct member & Variable or struct member is not assigned a value and then never used \\
        Unused private function & Private function is not called\\
        \bottomrule
        \end{tabular}
        
\label{fig:supressed_errors}
    \end{table}
During the initial analysis of these results, we noticed that many of the reported weaknesses were 
syntax errors, which are likely to be detected by most code editors and
eventually removed. Similar to Zhang et al., who ignored 129,395 instances of
syntax errors in their initial observation of 154,198 CWE instances, we ignore
such errors in our analysis. We also notice errors that were unfair to deem as a
weakness in relatively small code segments, such as CWE 563 - Assignment to
Variable without Use. These types of errors are not important as the intention
online Q/A sites is to address specific questions asked by the user, and not to
provide a complete solution. For example, in Listing ~\ref{563-ex} we see that
cppcheck detected an instance of CWE 563 - Assignment to Variable without Use in
line 7 since the variable \textit{ba} is not used within it's scope. Although
correct, this not a relevant error as it is evident from the comments that the
user posting the solution addressed the question asked, and left the completion
of the function up to the individual who asked the question. Therefore, we
cannot assume that the variable \textit{ba} will remain unused. As instances of
this nature are common, we suppress such errors in cppcheck by individually
selecting certain errors to be ignored in the final output. The suppressed
errors are summarized in Table ~\ref{fig:supressed_errors}.

\section{Empirical Study}\label{sec:study-results}
In this section, we report the results of an empirical study that we conducted by analyzing 
all the C/C++ IoT code examples found as weak in the five Stack Exchange sites: SO, Arduino, 
Raspberrry Pi, IoT, IoTa. We answer four research questions (RQ) are as follows.
\begin{enumerate}[label=\bf{RQ\arabic{*}}.]
  \item How many different types of weaknesses are found in the shared IoT code examples? (\sec\ref{sec:rq1})
  \item How do the observed IoT code weakness types map to CVE instances? (\sec\ref{sec:rq2})
  \item How are the mapped CVE types classified/categorized in the CVE details database? (\sec\ref{sec:rq3})
  \item How do the IoT weaknesses and vulnerabilities evolve over time? (\sec\ref{sec:rq4})
\end{enumerate}
\subsection{RQ 1 How many different types of weaknesses are found in the shared IoT code examples?}\label{sec:rq1}
\subsubsection{Motivation}
To gain a better understanding of the overall security of IoT related answers
found in the studied Stack Exchange sites, we analyze characteristics of the CWE
instances present in the obtained C and C++ snippets. Understanding which
specific CWE types and categories are more prevalent, as well as their
distribution among the three sites will provide insight on the nature of IoT
related vulnerabilities.
\subsubsection{Approach}
For each code snippet returned with one or more CWE ID by Cppcheck, we 
analyze the weakness type by consulting its description from the CWE database. 
We do this to understand the root cause behind the weakness for which the CWE is reported. 
For example, CWE 190 titled as ``Integer Overflow or Wraparound''. The description states \emt{The software performs a calculation that can produce an integer overflow or wraparound, when the logic assumes that the resulting value will always be larger than the original value. This can introduce other weaknesses when the calculation is used for resource management or execution control.} 
Thus, this weakness is due to the manipulation of memory location accessed through interger variable. 
Similarly, CWE 476 (NULL Pointer Derefence) is described as \emt{A NULL pointer dereference occurs when the application dereferences a pointer that it expects to be valid, but is NULL, typically causing a crash or exit.} 
This weakness is also caused due to the manipulation of a memory location via a C/C++ pointer. We thus 
group both CWE 190 and CWE 476 under a category `Memory'. 
We do this categorization for each of the distinct CWEs we find in our entire study dataset. 
The categorization is done by both the authors together, who consulted over Skype multiple times and revisited the categories several times 
to ensure the groups are unbiased and informative.  

\subsubsection{Results}

We found total 609 code examples, each of which was mapped to at least one known security weaknesses reported in the CWE databases. 
In total, the 609 code snippets showed 976 weaknesses.   
The breakdown of the number of weakness found on each individual site can be found in Table ~\ref{fig:studied_sites_weaknesses}.
As shown in the table, only snippets from Stack Overflow, Raspberry Pi, and
Arduino had vulnerabilities detected by Cppcheck, therefore these three sites
will be the focus of the rest of the study.  Among the five studied sites, we
did not find any code examples with a CWE map for code examples posted in the
two sites: IoT and IoTa. Among the other three sites, Arduino has the most
number of code snippets with security weaknesses, followed SO and Raspberry Pi.
    \begin{table}[t]
        \centering
        \caption{Weaknesses detected in each studied site}
        \begin{tabular}{lrr}\toprule
        \bf{Site Name} & \bf{\#Mapped CWEs (not distinct)} & \bf{\#Code Snippets with Mapped CWEs}\\
        \midrule
        Stack Overflow & 451 & 280\\
        Arduino & 465 & 296\\
        Raspberry Pi & 60  & 33\\
        IoT & 0 & 0\\
        IoTa & 0 & 0\\ 
        \bottomrule
        \end{tabular}
        
\label{fig:studied_sites_weaknesses}
    \end{table}

\begin{table}[t]
  \centering
  \caption{CWE types detected by Stack Exchange Site}
  \label{fig:total-weakness-results}
    \begin{tabular}{p{1cm}p{7cm}p{1cm}rll}
    \toprule{}
     \textbf{ID} & \textbf{CWE Title} & \textbf{\#CS} &\multicolumn{3}{c}{\bf{\%Distribution by Stack Exchange Site}} \\
 & & &SO    &ARD & RP\\
            \midrule
            398 & Code Quality & 422 &  48.3&46.7&5.0\\
            457 & Use of Uninitialized Variable & 42 & 33.3&59.5&7.2\\
            686 & Function Call With Incorrect Argument Type & 25 & 60.0&16.0&24.0\\
            571 & Expression is Always True & 24 & 25.0&70.8&4.2\\
            788 & Access of Memory Location After End of Buffer & 24 & 16.7&83.3&0\\
            595 & Comparison of Object References Instead of Object Contents & 23 & 39.1&60.9&0\\
            570 & Expression is Always False & 14 & 28.6&64.3&7.1\\
            758 & Reliance on Undefined, Unspecified Behavior & 14 & 71.4&21.4&7.1\\
            467 & Use of sizeof() on a Pointer Type & 8 & 62.5&25.0&12.5\\
            562 & Return of Stack Variable Address & 8 & 62.5&37.5&0\\
            561 & Dead Code & 8 & 12.5&75.0&12.5\\
            401 & Failure to Release Memory Before Removing Last Reference & 6 & 50.0&50.0&0\\
            783 & Operator Precedence Logic Error & 5 & 60.0&40.0&0\\
            190 & Integer Overflow or Wraparound & 5 & 40.0&20.0&40.0\\
            768 & Incorrect Short Circuit Evaluation & 4 & 25.0&75.0&0\\
            477 & Use of Obsolete Functions & 4 & 75.0&25.0&0\\
            685 & Function Call With Incorrect Number of Arguments & 4 & 50.0&0&50.0\\
            476 & NULL Pointer Dereference & 4 & 25.0&25.0&50\\
            252 & Unchecked Return Value & 3 & 66.7&33.3&0\\
            665 & Improper Initialization & 3 & 100&0&0\\
            682 & Incorrect Calculation & 2 & 0&100&0\\
            704 & Incorrect Type Conversion or Cast & 2 & 100&0&0\\
            664 & Improper Control of a Resource & 2 & 50.0&50.0&0\\
            369 & Divide By Zero & 1 & 0&100&0\\
            195 & Signed to Unsigned Conversion Error & 1 & 0&100&0\\
            628 & Function Call with Incorrectly Specified Arguments & 1 & 100&0&0\\
            683 & Function Call With Incorrect Order of Arguments & 1 &  0&100&0\\
            687 & Function Call With Incorrectly Specified Argument Value & 1 &  0&100&0\\
            672 & Operation on a Resource after Expiration or Release & 1 &  0&100&0\\
            \bottomrule
            \end{tabular}%
\end{table}

The 976 weakensses that we found belong to 29 distinct CWE types in the mitre CWE database. 
In the Mitre CWE database, 92 CWE types are related to C/C++. Therefore, we observe 32.2\% of all the 
listed C/C++ security weaknesses in the IoT code shared in the five Stack Exchange sites. 
In \tbl\ref{fig:total-weakness-results}, we show the distribution of the 29 CWE types in the three sites: SO, Arduino, and Raspberry Pi.
The first column is the ID of the CWE type, the second column shows its title, the third column (\#CS) shows the total number 
of code snippets found with the CWE type across all the three sites. The last column (`\% Distribution Across Sites') shows the percent distribution of those 
code snippets per the three sites. The CWE types are sorted based on the \#CS. The most frequent 
weakness observed was CWE-398 (Code Quality). This CWE type considers any poor coding practice as a weakness in the code, whether or not 
that may result any security concern. This weakness was found total 422 times (i.e., in 69.2\% of all code snippets labeled as weak by Cppcheck).
Listing ~\ref{398-ex} is example of a code snippet that
cppcheck identified having an occurrence of CWE 398 in line 15. This is due to
variable shadowing of the variable "lastSwitchOneState",  which was already
declared outside the function block in line 1. Although this is not a direct
vulnerability, it is a coding practice that may lead to unpredictable behaviour
and thus is poor quality.
\begin{figure}[ht]
\begin{lstlisting}[language=C,label=398-ex,caption=CWE 398 - Code Quality in an Arduino Code Snippet]
int lastSwitchOneState = 0;     // previous state of the switch

int switchTwoState = 0;
int lastSwitchTwoState = 0;

int switchThreeState = 0;
int lastSwitchThreeState = 0;


void setup() {
    //initialize serial communication at 9600 bits per second:
    Serial.begin(9600);

    int switchOneState = 0;         // current state of the switch
    int lastSwitchOneState = 0;     // previous state of the switch switch pins as an input
    pinMode(switchOnePin, INPUT);
    pinMode(switchTwoPin, INPUT);
    pinMode(switchThreePin, INPUT);
}
\end{lstlisting}
\end{figure} 

Among the other 28 distinct CWE types that were found, we observe that CWE 457 - Use of
Uninitialized Variable was the most frequently detected CWE type and was present
in 42 out of the 240 code snippets (17.5\%). In the rest of the paper, we focus our analysis on the rest of non-CWE-398 weaknesses, given 
those weaknesses may be more severe/likely to introduce security concern.
 
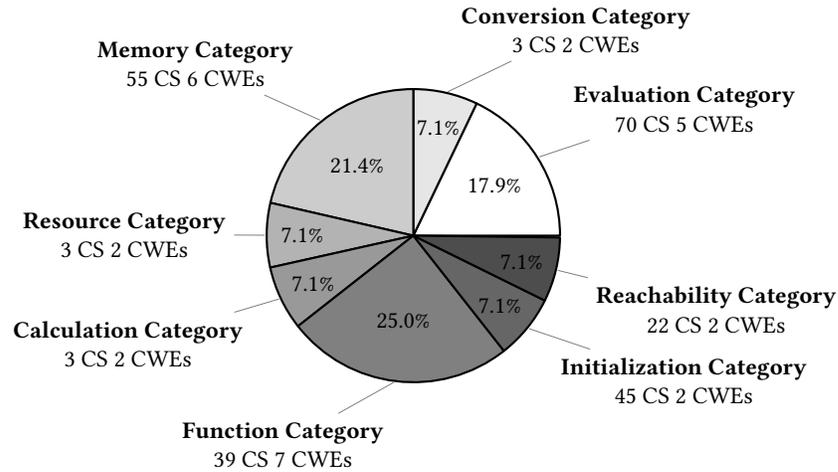
\begin{figure}[ht]
    \centering\begin{tikzpicture}[scale=.65]-
    \pie[
       /tikz/every pin/.style={align=center},
        text=pin,
        explode=0.0, 
        color={black!0, black!10, black!20, black!30, black!40, black!50, black!60, black!70}
        ]
        {
            17.9/\bf{Evaluation Category}\\70 CS 5 CWEs,
            7.1/\bf{Conversion Category}\\3 CS 2 CWEs,
            21.4/\bf{Memory Category}\\55 CS 6 CWEs,
            7.1/\bf{Resource Category}\\3 CS 2 CWEs,
             7.1/\bf{Calculation Category}\\3 CS 2 CWEs,
            25.0/\bf{Function Category}\\39 CS 7 CWEs,
            7.1/\bf{Initialization Category}\\45 CS 2 CWEs,
            7.1/\bf{Reachability Category}\\22 CS 2 CWEs
        }
    \end{tikzpicture}
	\caption{Distribution of the observed 28 CWE instances (i.e., non-CWE-398) by the eight weakness categories in our studied dataset across the three sites 
	(CS = weak code snippet)}
	\label{fig:vul-cat-dist}
\end{figure}After excluding instances of CWE 398, there are a total of 348
instances of weaknesses in 240 code snippets. From these distinct CWE instances,
we group them into the following 8 weakness categories based on common
characteristics: Evaluation, Memory, Function, Initialization, Reachability,
Resource, Conversion, and Calculation. As shown in Figure
~\ref{fig:vul-cat-dist}, the Evaluation category had the greatest number of
vulnerable code snippets (70 out of 240 CS, or 29.17\% ), while weaknesses
related to Conversion errors were only present in 3 total code snippets
(1.25\%). This suggests that users on Stack Exchange sites commonly introduce
errors that contain improper evaluation methods rather than improper
conversions. We also observe that the weakness categories containing the
greatest number of individual CWE types are the Function and Memory categories,
with 7 and 6 CWE types respectively. 

Among the three studied sites, we find that the distribution of the 8 general
weakness categories varies depending on the specific category. As shown in
\tbl\ref{fig:cwe_cat_distribution}, half of the weakness categories (4/8)
occur more frequently in Arduino code snippets, followed by Stack Overflow and
Raspberry Pi. However, we also find that a few of the categories, such as
Function and Conversion, are more commonly found in Stack Overflow code
snippets. In the case of the Function category, the difference is significant as
59.0\% of code snippets with function related weaknesses were obtained from
Stack Overflow. This is likely due to most Stack Overflow code snippets
containing C++ code, unlike the analyzed Arduino and Raspberry Pi code snippets
which are entirely C code.
\begin{table}[t]
  \centering
  \caption{Distribution of the observed 28 CWE Categories (i.e., non-CWE-398) in our dataset by categories and by Stack Exchange site, CS = Weak code snippet}
  \label{fig:cwe_cat_distribution}
    \begin{tabular}{p{5cm}p{3cm}p{2cm}p{2cm}p{2cm}}
    \toprule{}
    \textbf{CWE Category} & \textbf{CS}&  \multicolumn{3}{c}{\bf{\%Distribution of CS by Stack Exchange Site}}\\
 & &SO    &ARD & RP\\
    \midrule
    Evaluation& 70 & 32.9&64.3&2.8\\
    Memory& 55 & 38.2&52.6&9.2	\\
    Function& 39 & 59.0&20.5&20.5\\
    Initialization& 45 & 37.8&55.6&6.6\\
    Reachability& 22 & 50.0&40.9&9.1\\
    Resource& 3 & 33.3&66.7&0\\
    Conversion& 3 & 66.7&33.3&0\\
    Calculation& 3 & 0&100&0\\
    
    \bottomrule
    \end{tabular}%
\end{table}

The 8 general weakness categories are broken down below where we further analyze
their CWE instances. We group these instances by their parent CWE category type
as found on cwe.mitre.org. Overall, we observe a total of 7 official parent CWE
categories, while 8 out of the 28 CWE types do not belong to any parent category.

\nd\bf{$\bullet$ Function Type Weaknesses - 7 CWE (16.3\% of Weak Code) }

\begin{figure}[ht]
\centering
\includegraphics[scale=0.6]{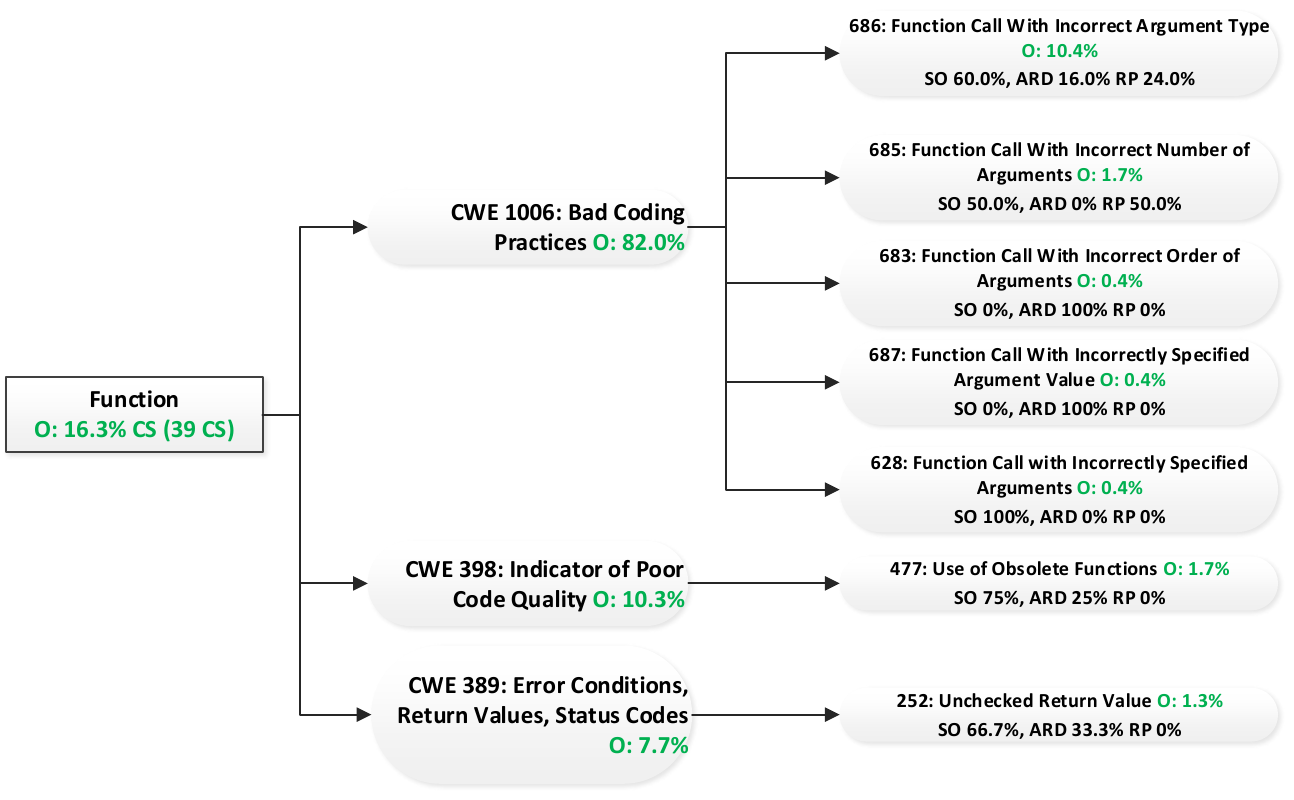}
\caption{Function type with CWE categories and types (SO = Stack Overflow, AD = Arduino, RP = Raspberry Pi, O = Overall). Overall \% of CWE types based on total number of weak code snippets, i.e. 240}
\label{fig:fn_flowchart}
\end{figure}

The Function category contains weaknesses that involve the incorrect use or design of functions. As shown in Figure ~\ref{fig:fn_flowchart}, 7 out of the distinct 28 CWE types are categorized as this type. These CWE types belong to 3 different CWE categories, with the majority of vulnerable code snippets belonging to CWE 1006-Bad coding practices (82.0\%). We find that the most frequently detected CWE type is CWE 686 - Function Call With Incorrect Argument Type, and is present in 25 code snippets with weaknesses. We also find that the majority of the weaknesses related to functions occur in StackOverflow code snippets. This is likely due to more C++ code snippets obtained from StackOverflow. In Section \ref{sec:rq3} where we analyze CVE instances of CWEs in the function category, we observe that some are related to Denial of Service and code execution. When vulnerabilities like this occur in real work software systems, users may be unable to access services and potentially important personal information.

An example of a function related weakness that was detected in StackOverflow code snippet is shown in Listing ~\ref{fn-ex}, which depicts an instance of CWE 686 - Function Call With Incorrect Argument Type. We see that the argument type the printf function is expecting is "int", however it is receiving the return value of the sizeof function, which is of type "size\_t".
\begin{figure}[ht]
\begin{lstlisting}[language=C,label=fn-ex,caption=CWE 686 - Function Call With Incorrect Argument Type in a StackOverflow code snippet]
int main(void) {
        printf("int has %d bytes\n",sizeof(int));
        printf("long has %d bytes\n",sizeof(long));
\end{lstlisting}
\end{figure} 

\nd\bf{$\bullet$ Memory Type Weaknesses - 6 CWE (22.9\% of Weak Code)}

\begin{figure}[ht]
\centering
\includegraphics[scale=0.6]{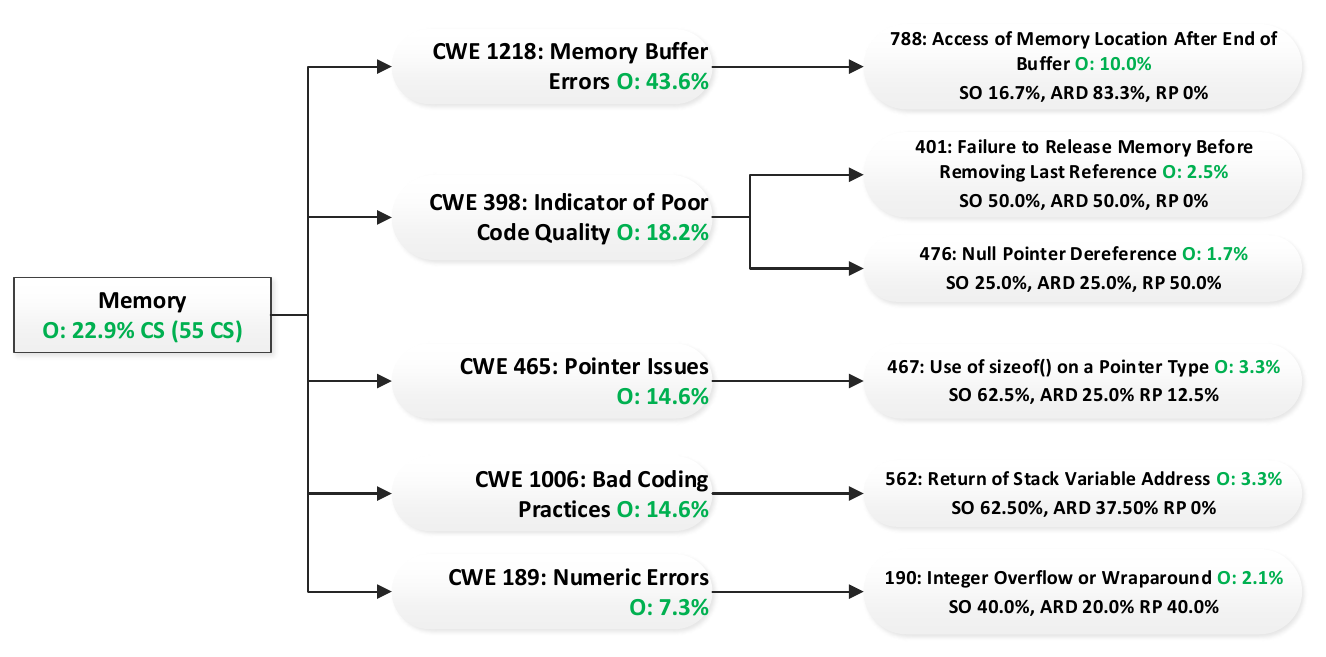}
\caption{Memory type with CWE categories and types(SO = Stack Overflow, AD = Arduino, RP = Raspberry Pi, O = Overall)}
\label{fig:me_flowchart}
\end{figure}

\nd The Memory type contains weaknesses involving operations that mismanage a program's memory. Figure ~\ref{fig:me_flowchart} shows that within the obtained vulnerable code snippets, there are 6 CWE types that can be categorized as this type. These vulnerabilities belong to 4 different CWE categories, with the majority belonging to CWE 1218-Memory Buffer Errors (50\%). The most frequently detected Memory related CWE type is CWE 788 - Access of Memory Location After End of Buffer, and is present in 24 weak code snippets. Among the three studied sites, we find that most of the CWE types occur more frequently in StackOverflow code snippets (4/6). When we analyze CVE instances of these CWE types in Section \ref{sec:rq3}, we observe that some are related to Buffer overflow, which can allow attackers to make changes to the memory of a software system, and potentially expose private user information.
\newline
An example of a memory weakness that was detected in an Arduino code snippet is shown in listing ~\ref{me-ex}, which depicts an instance of CWE 788 - Access of Memory Location After End of Buffer. We see that in line 19, fifo\_bounds[8] is an access after buffer size as in line 3 it was declared with a size of 8. 

\begin{figure}[ht]
\begin{lstlisting}[language=C,label=me-ex,caption=CWE 788-Access of Memory Location After End of Buffer in an Arduino code snippet]
static unsigned long timeout = 0; // used to detect dead sensor/connection
uint16_t fifo_count;
uint8_t fifo_buffer[8];
int16_t acc_temp[3];
int16_t gyro_x;

void loop(){

  if(state_fifo_int){ // interrupt has triggered
    blink(0,1);  // really quick blink // really quick blink
    //Serial.println(x);
    state_fifo_int = false;
    accelGyroMag.resetFIFO();
    accelGyroMag.getFIFOBytes(fifo_buffer,8); // fill in fifo bytes
    accelGyroMag.getIntFIFOBufferOverflowStatus(); // read the INT_STATUS reg in order to clear the Latch. 
    acc_temp[0] = (((int16_t)fifo_buffer[0]) << 8) | fifo_buffer[1];
    acc_temp[1] = (((int16_t)fifo_buffer[2]) << 8) | fifo_buffer[3];
    acc_temp[2] = (((int16_t)fifo_buffer[4]) << 8) | fifo_buffer[5];
    gyro_x = (((int16_t)fifo_buffer[7]) << 8) | fifo_buffer[8];

    timeout = millis();

  }
.....
}
\end{lstlisting}
\end{figure} 

\nd\bf{$\bullet$ Evaluation Type Weaknesses - 5 CWE (29.2\% of Weak Code)}

\begin{figure}[ht]
\centering
\includegraphics[scale=0.50]{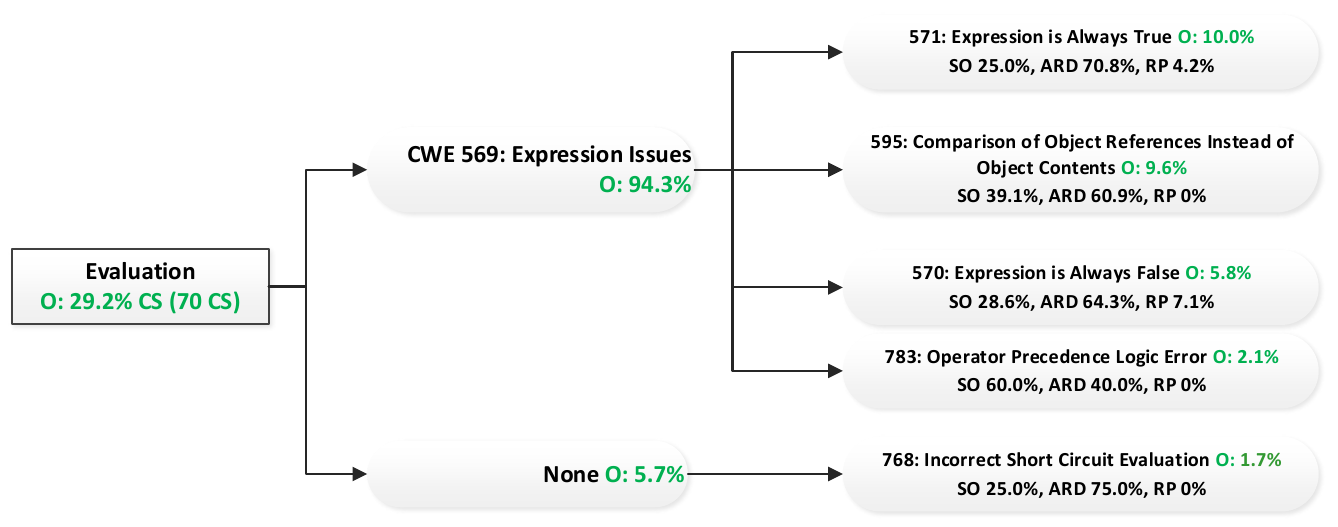}
\caption{Evaluation type with CWE categories and types (SO = Stack Overflow, AD = Arduino, RP = Raspberry Pi, O = Overall)}
\label{fig:ev_flowchart}
\end{figure}

\nd The Evaluation type contains weaknesses that involve incorrect evaluations such as improper comparisons or logic errors. As shown in Figure ~\ref{fig:ev_flowchart}, 5 out of the 28 distinct CWE types are categorized as an Evaluation weakness, with 4 belonging to CWE category 569-Expression Issues (94.17\%). We find that CWE 571 - Expression is always true is the most frequently occurring CWE type and is present in 24 weak code snippets. When looking at the distribution of Evaluation type vulnerabilities among the three sites, we see that they are more prevalent in Arduino code snippets for 4 out of the 5 CWE types.
\newline
Listing ~\ref{ev-ex} shows an example of a Raspberry Pi code snippets that an evaluation weakness in line 3, where cppcheck detected CWE 570 - Expression is Always False due to identical statements in the OR operation which may result in that expression always returning as false. 
\begin{figure}[ht]
\begin{lstlisting}[language=C, label=ev-ex,caption=CWE 570-Expression is Always False in a RaspberryPi code snippet]
char * rstrip(char *s) {
    char *p = s + strlen(s)-1;
    while((*p == '\n')||(*p == '\n')) {
        p--;
    }
    *p = '\0';
    return s;
}
\end{lstlisting}
\end{figure}

\nd\bf{$\bullet$ Initialization Type Weaknesses - 2 CWE (18.8\% of Weak Code)}

\begin{figure}[ht]
\centering
\includegraphics[scale=0.55]{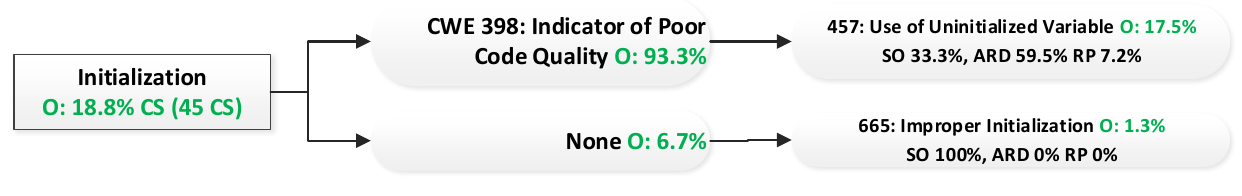}
\caption{Initialization type with CWE categories and types (SO = Stack Overflow, AD = Arduino, RP = Raspberry Pi, O = Overall)}
\label{fig:in_flowchart}
\end{figure}

\nd The Initialization type contains weaknesses that involve either improperly initializing a variable, or using an uninitialized variable. Figure ~\ref{fig:in_flowchart} shows that there are 2 CWE types that belong to this type, with CWE 457- Use of Uninitialized Variable being present in the greatest number of code snippets (42). It also belongs to the only CWE category in this weakness type; CWE 398 - Indicator of Poor Code Quality. Although we did not observe a large number of initialization related CWE types, they were still present in a large number of the code snippets. We also see that the distribution code snippets containing initialization weaknesses among the three sites differs between the two CWE types. For CWE 457, each site contains instances of this type with the majority detected in Arduino code snippets. However, for CWE 665- Improper Initialization, there are only instances detected in Stack Overflow code snippets. This is likely due to only 3 total code snippets containing CWE 665. 
In Section \ref{sec:rq3} where we analyze CVE instances of these CWE types, we find that some are related to bypassing, which allows hackers to gain access to information normally hidden behind authentication mechanisms.
\newline
An example of a initialization weakness (CWE 457) that was detected in an Arduino code snippet is shown in listing ~\ref{in-ex}, where we can see that in line 2 there was a failure to initialize the x variable in the for loop.
\begin{figure}[ht]
\begin{lstlisting}[language=C,label=in-ex,caption=CWE 457- Use of Uninitialized Variable in an Arduino code snippet]
  if (cnt == 0){
    for(int x; x < 8; x++){
      led[x] = random(2);
    }
  }
\end{lstlisting}
\end{figure}

\nd\bf{$\bullet$ Reachability Type Weaknesses - 2 CWE (9.2\% of Weak Code)}

\begin{figure}[ht]
\centering
\includegraphics[scale=0.55]{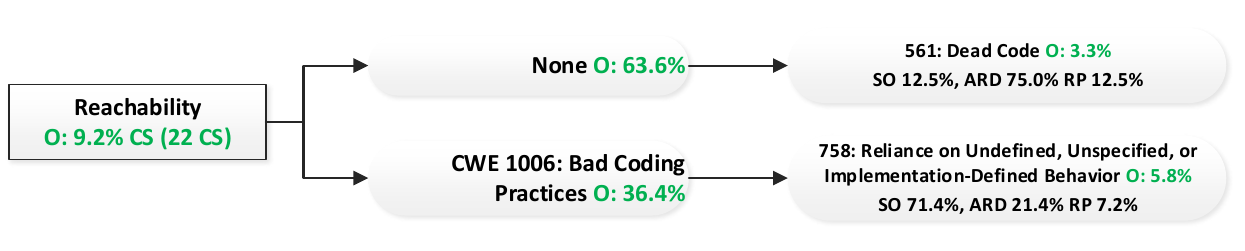}
\caption{Reachability type with CWE categories and types (SO = Stack Overflow, AD = Arduino, RP = Raspberry Pi, O = Overall)}
\label{fig:rc_flowchart}
\end{figure}

\nd The Reachability type contains weaknesses that involve unreachable or undefined code which may lead to unpredictable results. As shown in Figure ~\ref{fig:rc_flowchart}, we find that 2 out of the distinct 28 CWE types are categorized as this type. Only one CWE category belongs to this general vulnerability type (CWE 1006-Bad coding practices), while the majority of the CWE instances related to reachability do not belong to any CWE category (63.6\%). CWE 758-Reliance on Undefined, Unspecified Behavior is the most frequently occurring CWE type and is present in 14 weak code snippets. After we analyze the distribution of these CWE instances among the three sites, we find that for CWE 561 - Dead Code , most of the instances occur in Arduino snippets (75.0\%), while for CWE 758 71.4\% of the instances occur in Stack Overflow snippets.
\newline
Listing ~\ref{rc-ex} shows an example of a reachability weakness (CWE 561-Dead Code) that was detected in an Arduino code snippet. The "dead-code" statement occurs in line 12 as this break statement will never be executed and the program will never reach that line.
\begin{figure}[ht]
\begin{lstlisting}[language=C,label=rc-ex,caption=CWE 561-Dead Code in an Arduino code snippet]
  if (ok.wasReleased()) {       
            switch (menu_current) {          
                case 0: 
                    screen = 2;
                    break;
                case 1:
                    screen = 3;
                    break;
                case 2:
                    screen = 0;
                    break;
                break;
            }
\end{lstlisting}
\end{figure}

\nd\bf{$\bullet$ Resource Type Weaknesses - 2 CWE (1.3\% of Weak Code)}

\begin{figure}[ht]
\centering
\includegraphics[scale=0.55]{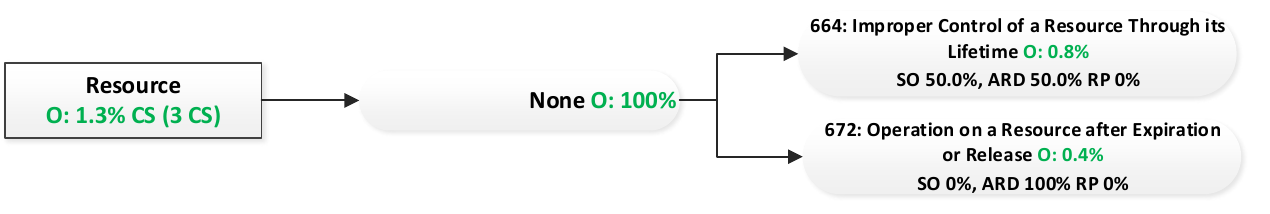}
\caption{Resource type with CWE categories and types (SO = Stack Overflow, AD = Arduino, RP = Raspberry Pi, O = Overall)}
\label{fig:re_flowchart}
\end{figure}

\nd The Resource type contains weaknesses that involve code mismanagement of a program's resources. Figure ~\ref{fig:re_flowchart} shows that 2 CWE types are related to Resource weaknesses, and both types do not belong to any CWE category. From these two types, CWE 664-Improper Control of a Resource Through its Lifetime is the most frequent (2 weak code snippets). We also find that two the CWE types are similar in that they have zero occurrences in Raspberry Pi code snippets. When we analyze the CVE instances of these CWE types in Section \ref{sec:rq3} , we find that some are we observe that some are related to code execution attacks such as arbitrary code execution, which allows attackers to execute harmful code and potentially comprise user information.
\newline
An example of incorrect handling of a resource (CWE 664) that was detected in an Arduino code snippet is shown in listing ~\ref{re-ex}, where we can see that the "va\_list" variable was opened but not closed by a "va\_end()" statement before the end of the function.
\begin{figure}[ht]
\begin{lstlisting}[language=C,label=re-ex,caption=CWE 664-Improper Control of a Resource Through its Lifetime in an Arduino code snippet]
int ardprintf(char *str, ...)
{
  int i, count=0, j=0, flag=0;
  char temp[ARDBUFFER+1];
  for(i=0; str[i]!='\0';i++)  if(str[i]=='%')  count++;

  va_list argv;
  va_start(argv, count);
  for(i=0,j=0; str[i]!='\0';i++)
  {
    if(str[i]=='%')
    {
      temp[j] = '\0';
      Serial.print(temp);
      j=0;
      temp[0] = '\0';

      switch(str[++i])
      {
        case 'd': Serial.print(va_arg(argv, int));
                  break;
        case 'l': Serial.print(va_arg(argv, long));
                  break;
        case 'f': Serial.print(va_arg(argv, double));
                  break;
        case 'c': Serial.print((char)va_arg(argv, int));
                  break;
        case 's': Serial.print(va_arg(argv, char *));
                  break;
        default:  ;
      };
    }
    else 
    {
      temp[j] = str[i];
      j = (j+1)%ARDBUFFER;
      if(j==0) 
      {
        temp[ARDBUFFER] = '\0';
        Serial.print(temp);
        temp[0]='\0';
      }
    }
  };
  Serial.println();
  return count + 1;

}
\end{lstlisting}
\end{figure}

\nd\bf{$\bullet$ Conversion Type Weaknesses - 2 CWE (1.25\% of Weak Code)}

\begin{figure}[ht]
\centering
\includegraphics[scale=0.55]{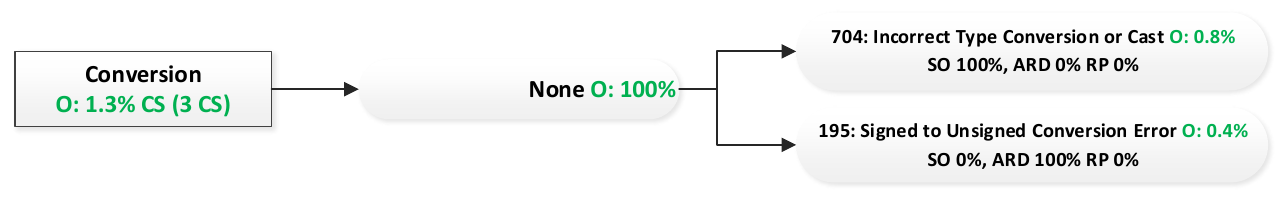}
\caption{Conversion type with CWE categories and types (SO = Stack Overflow, AD = Arduino, RP = Raspberry Pi, O = Overall)}
\label{fig:cn_flowchart}
\end{figure}

\nd The Conversion type contains weakness related to incorrect conversions of variables to different types. As shown in Figure ~\ref{fig:cn_flowchart}, we find 2 CWE types that are related to conversion errors with both not belonging to any CWE category. Out of the two types, CWE 704-Incorrect Type Conversion or Cast contains is the most frequent among weak code snippets (2). We also find that both types do not have any instances in Raspberry Pi code snippets. Similar to Resource weaknesses, we observe that CVE instances of these CWE types analyzed in Section \ref{sec:rq3} are related to code execution attacks such as arbitrary code execution. 
\newline
Listing ~\ref{cn-ex} shows an example of a conversion error in a Stack Overflow code snippet. We see that line 4 is an occurrence of CWE 704 as there is a casting between an unsigned long pointer and a float pointer.  
\begin{figure}[ht]
\begin{lstlisting}[language=C,label=cn-ex,caption=CWE 704-Incorrect Type Conversion or Cast in a Stack Overflow code snippet]
unsigned long d;
d =  (outbox[index+3] << 24) | (outbox[index+2] << 16)
        | (outbox[index+1] << 8) | (outbox[index]);
float member = *(float *)&d;
\end{lstlisting}
\end{figure}

\nd\bf{$\bullet$ Calculation Type Weaknesses - 2 CWE (1.25\% of Weak Code)}

\nd The Calculation type contains the least number of CWE instances among the 8 general vulnerability types. Weaknesses that are part of this type involve improper or incorrect calculations. As shown in Figure ~\ref{fig:fn_flowchart}, we find that this general vulnerability type contains 2 CWE types and one CWE category (CWE 189 Numeric Errors). CWE 682- Incorrect Calculation is present in the greatest number of weak code snippets (2) and does not belong to any CWE category. We also find that all of the calculation related weakness were only detected in Arduino code snippets. Furthermore, we find in Section \ref{sec:rq3} that CVEs related to these CWE types can be labelled as "Gain Information" and "Gain Privileges" vulnerabilities. This indicates that when calculation type errors occur in real world software, they have the potential to compromise information and access. 
\begin{figure}[ht]
\centering
\includegraphics[scale=0.6]{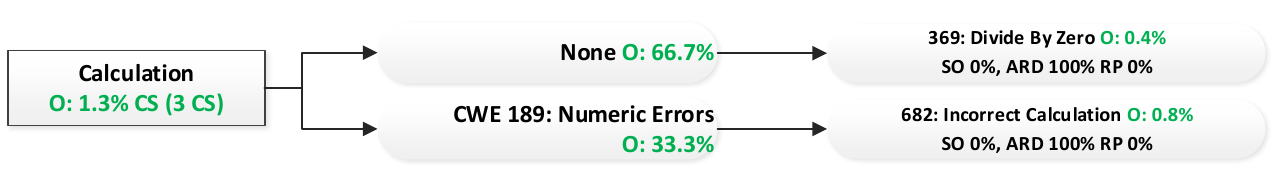}
\caption{Calculation type with CWE categories and types (SO = Stack Overflow, AD = Arduino, RP = Raspberry Pi, O = Overall)}
\label{fig:ca_flowchart}
\end{figure}
\newline
An example of a calculation vulnerability (CWE 682- Incorrect Calculation) that was detected in an Arduino code snippet is shown in listing ~\ref{ca-ex}. We can see that in line 2, the error made was there is a calculation inside of the sizeof function. 
\begin{figure}[ht]
\begin{lstlisting}[language=C,label=ca-ex,caption=CWE 682- Incorrect Calculation in an Arduino code snippet]
  // Read 12 bytes (11 bytes of data + 1 byte delimiter)
  int nBytes = Serial.readBytes(packet, sizeof(CHANNELS + 1));
\end{lstlisting}
\end{figure}

\begin{tcolorbox}[flushleft upper, boxrule=1pt, arc=0pt, left=0pt, right=0pt, top=0pt, bottom=0pt, colback=white, after=\ignorespacesafterend\par\noindent]
\textbf{Summary of RQ1: Types of CWE Weaknesses in Vulnerable IoT C/C++ Code Snippets} 
We find 29 out of 90 C/C++ CWE types in the Mitre database in three of our studied sties: SO, Arduino, and Raspberry Pi. 
The majority (69\%) of the code snippets with the weakensses belong to CWE-398 (Poor Code Quality). 
The rest of the 28 CWEs are grouped into 8 weakness categories. 
We observe that the most weaknesses occur due to Function, Memory, and Evaluation related errors, which contain 7, 6, and 5 CWE types respectively. 
Among the three sites, Arduino has the most number of IoT C/C++ shared code snippets with weaknesses, followed by SO.
\end{tcolorbox}

\subsection{RQ 2 How do the observed IoT code weakness types map to CVE instances?}\label{sec:rq2}
\subsubsection{Motivation}
The CWE weakness types in Mitre database are documented, because the weaknesses can be subject to potential 
exploitation by malicious users when found in real-world IoT software systems. Therefore, it would 
be important to know whether and how the CWE types we observed in our IoT dataset were exploited to create 
real-world software vulnerabilities. The CVE (Common Vulnerability Exposure) entries 
offer information about such vulnerabilities. An analysis of of how our observed CWE types map to the CVE instances 
may guide our efforts like prioritization of which weakness types need to be addressed first.  
\subsubsection{Approach}\label{sec:rq2-approach}
We first use cvedetails.com to collect the CVE instances for each detected CWE
type in the code snippets. After determining the total numbers of instances for
each type, we then observe how they are distributed within the 8 general
weakness categories as analyzed in RQ1. A CWE entry can map to zero, one, or more than one 
CVE instances. We analyze the CVE instances per CWE category from RQ1 as follows:
\begin{itemize}
  \item We find each CWE instance name with its distribution as total number of CVEs mapped, total code snippets belonging to the CWE in our dataaset. 
  \item For each CWE instance, we discuss the most severe CVE instances that are mapped to it. 
  We assess severity of a CVE instance based on the CVSS score of the instance as found in the cvedetails.com database. A CVSS score \cite{cvss} can range between 0-10 (10 being the most severe). To determine severity, we use the CVSS score given to each instance. We refer to the CVSS v3.0 ratings which groups the scores into the following severity types: Low (0.1-3.9), Medium (4.0-6.9), High (7.0-8.9), Critical (9.0-10.0).
  
  \item For each CWE instance, we create a wordcloud by taking as input the description of all mapped CVEs. 
\end{itemize}

\subsubsection{Results}
\begin{table}[t]
  \centering
  \caption{Distribution  of the 12 CWEs with mapped CVEs in the cvedetails.com database by CVSS severity category as defined in \sec\ref{sec:rq2-approach}}
  \label{fig:cvss_distribution_by_CWE}
    \begin{tabular}{p{8cm}p{1cm}p{2cm}lll}
    \toprule{}
    \textbf{CWE Category} & \bf{\#CVEs} &\multicolumn{4}{c}{\bf{\%Distribution by CVSS Score Category}}\\
 & &Low    & Medium &  High & Critical \\
    \midrule
    \textbf{Memory Category} & 2997 & 5.6&72.3&16.1&6.0\\
    CWE-190 Integer overflow/wraparound & 1418	& 3.1&71.0&18.4&7.6\\
    CWE-476 Null pointer dereference & 1371 & 7.6&74.1&13.4&5.0	\\
    CWE-401	Failure to release memory & 195 & 10.8&69.7&19.5&0\\
    CWE-788 Access memory location	& 13 & 0&53.9&7.7&38.5\\
    \midrule
    {\textbf{Calculation Category}} & 239 & 14.2&74.5&9.2&2.1\\
    CWE-369 Divide by zero  & 196	& 16.3&77.6&6.1&0\\
    CWE-682 Incorrect calculation & 43 & 4.7&60.5&23.2&11.6\\
    \midrule 
    {\textbf{Conversion Category}} &165 & 1.2&78.8&6.1&13.9\\
    CWE-704 Incorrect type conversion & 165	& 1.2&78.8&6.1&13.9\\
    \midrule
    {\textbf{Initialization Category}} & 148 & 35.8&46.6&10.2&7.4 \\
    CWE-665 Improper initialization & 148	& 35.8&46.6&10.2&7.4 \\
    \midrule
    {\textbf{Function Category}} & 31 & 6.5&61.3&25.8&6.4\\
    CWE-252 Unchecked return value & 31	& 6.5&61.3&25.8&6.4\\
    \midrule
    {\textbf{Resource Category}} & 14 & 7.1&35.7&42.9&14.3\\
    CWE-672 Resource operation after release & 12	& 8.3&33.3&50.0&8.3		\\
    CWE-664 Improper control of resource & 2 & 0&50.0&0&50.0\\
    \midrule
    \bf{Overall} & 3595& 7.3&71.4&15.1&6.2\\
    \bottomrule
    \end{tabular}%
    \label{tab:mapped-cwe-cve}
\end{table}%
\begin{figure}[t]
    \centering
    \subfloat[CWE 190]{
        \label{190wordcloud}
        \includegraphics[width=0.3\textwidth]{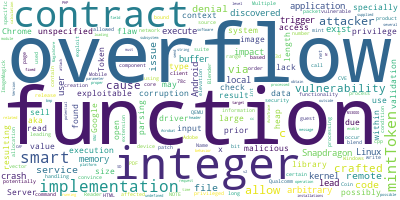}
    }
    \subfloat[CWE 476]{
        \label{476wordcloud}
        \includegraphics[width=0.3\textwidth]{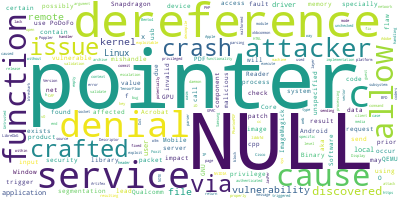}
    }
    \subfloat[CWE 401]{
        \label{401wordcloud}
        \includegraphics[width=0.3\textwidth]{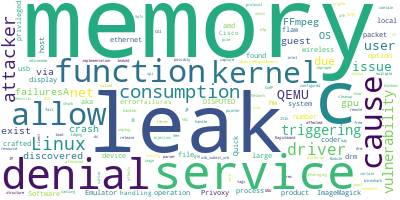}
    }
    \newline
     \subfloat[CWE 788]{
        \label{788wordcloud}
        \includegraphics[width=0.3\textwidth]{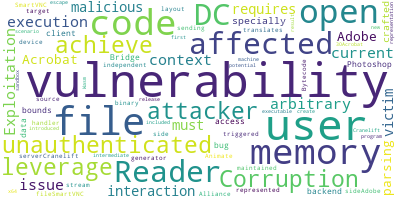}
    }
    \subfloat[CWE 369]{
        \label{369wordcloud}
        \includegraphics[width=0.3\textwidth]{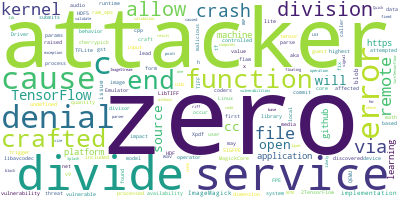}
    }
    \subfloat[CWE 682]{
        \label{682wordcloud}
        \includegraphics[width=0.3\textwidth]{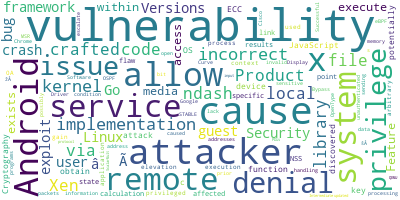}
    }
    \newline
     \subfloat[CWE 704]{
        \label{704wordcloud}
        \includegraphics[width=0.3\textwidth]{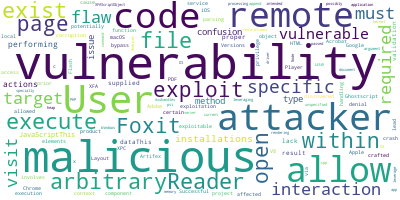}
    }
    \subfloat[CWE 665]{
        \label{665wordcloud}
        \includegraphics[width=0.3\textwidth]{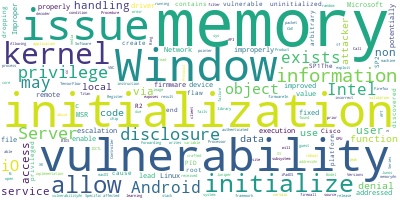}
    }
    \subfloat[CWE 457]{
        \label{457wordcloud}
        \includegraphics[width=0.3\textwidth]{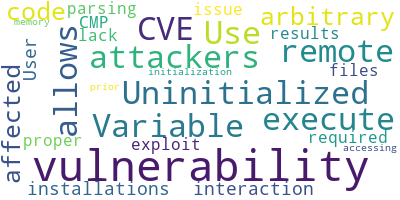}
    }
    \newline
     \subfloat[CWE 252]{
        \label{252wordcloud}
        \includegraphics[width=0.3\textwidth]{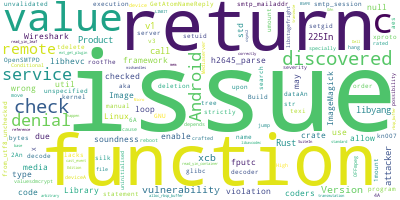}
    }
    \subfloat[CWE 672]{
        \label{672wordcloud}
        \includegraphics[width=0.3\textwidth]{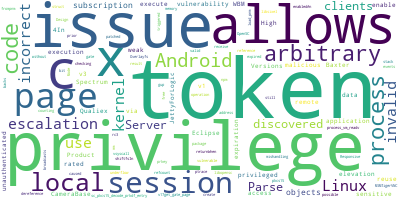}
    }
    \subfloat[CWE 664]{
        \label{664wordcloud}
        \includegraphics[width=0.3\textwidth]{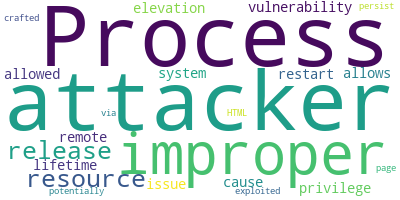}
    }
    \caption{Frequently occurring keywords in descriptions of highest severity CVE instances of each CWE type}
    \label{fig:overall_wordcloud}
\end{figure}
\begin{table}[t]
        \centering
        \caption{Highest severity CVE instances of each CWE type with CVE instances}

        \begin{tabular}{p{3cm}lrp{8cm}}
        \toprule
        \bf{CWE Type} & \bf{CVE ID} & \bf{CVSS} & \bf{Description}\\
	\midrule
	190 - Integer Overflow & 2020-11279 & 10 & Memory corruption due to improper length check in SDES packets\\
	476 - NULL Pointer Derefence & 2020-11168 & 10 & While accessing data buffer beyond its size.\\
	401 - Failure to Release Memory  & 2019-19069 & 7.8 & A memory leak in the fastrpc\_dma\_buf\_attach() function in the Linux kernel allows attackers to cause a denial of service (memory consumption).\\
	788 - Access of memory after buffer & 2021-21048 & 9.3 & Memory Corruption vulnerability when parsing a specially crafted file in Adobe. Exploitation of this issue requires a victim must open a malicious file.\\
	  
	369 - Divide by Zero & 2012-0207 & 7.8 & The igmp\_heard\_query function in the Linux kernel allows remote attackers to cause a denial of service (divide-by-zero error and panic) via IGMP packets.\\
	682 - Incorrect Calc & 2017-13151 & 9.3 & A remote code execution vulnerability in the Android media framework (libmpeg2).\\
	704 - Incorrect Type Conversion & 2018-15981 & 10 & Flash Player versions have a type confusion vulnerability. Successful exploitation could lead to arbitrary code execution.\\
	665 - Improper Initialization & 2018-11949 & 10 & Failure to initialize the extra buffer leads to an out of buffer access.\\
	457 -  Use of Uninitialized Variable & 2021-31435 & 6.8 & This vulnerability allows remote attackers to execute arbitrary code on affected installations of Foxit Studio Photo.\\
	252 - Unchecked return value & 2017-0720 & 9.3 & A remote code execution vulnerability in the Android media framework (libhevc). \\
	672 - Resource after Expiration/Release & 2017-0544 & 9.3 &An elevation of privilege vulnerability in CameraBase could enable a local malicious application to execute arbitrary code. \\
	664 - Improper Control of a Resource & 2016-8763 & 9.3 &This vulnerability in the TrustZone driver allows attacker to cause system restart or privilege escalation.
	\\
	        \bottomrule
        \end{tabular}
    
    \label{fig:highest-severity}
\end{table}
We observe a total of 3595 CVE instances in 12 out of the 29 distinct CWE types
identified in our dataset (from RQ1). The 12 CWE types belong to 6 out of the 8
general weakness categories we determine in RQ1 (see \tbl\ref{tab:mapped-cwe-cve}).  
The second column in  \tbl\ref{tab:mapped-cwe-cve} (\#Mapped CVEs) shows the total number of CVEs mapped to each of the 12 CWEs. 
The last column (CVSS score) shows the percent distribution of CVSS scores across the mapped CVEs along four severity groups as defined in \sec\ref{sec:rq2-approach}.
For example, CWE-190 is found in 1418 CVEs. Out of the mapped CVEs under CWE-190, 3.1\% are of low severity, 
71\% are of medium, 18.4\% are of high, and 7.6\% are of critical severity.   

As shown in \tbl\ref{tab:mapped-cwe-cve}, the vast majority of the CVE instances belong to
CWE types that can be categorized as \textit{Memory} related weakeness category. We
also observe this in \tbl\ref{tab:mapped-cwe-cve} that CWE
190 - Integer Overflow or Wraparound and CWE 476 - NULL Pointer Dereference have
the greatest number of CVE instances (1418 and 1371 respectively), and are both
categorized under the \textit{Memory} type. Overall, out of the CWE eight weakness categories, we find that the mapped 12 CWEs belong to six categories: 
Memory, Calculation, Conversion, Initialization, Function, and Resource. We did not find any CVEs mapped to the CWEs from two weakness categories: Reachability and Evaluation. 
Although we find that
16.3\% of all weak code snippets are contain are \textit{Function} type
weakness, only 0.9\% of the total CVE instances (31 out of 3595) are from CWE
types belonging to the \textit{Function} category. This finding may offer more confidence to the code examples shared in the Stack Exchange 
sites, because most of the CWEs (17) do not have any CVEs mapped. As shown in \tbl\ref{tab:mapped-cwe-cve}, most the mapped 
CVEs belong to Medium severity (71.4\%), followed High severity (15.1\%), Low (7.3\%), and Critical severity (6.2\%).  

One possible explanation as to why certain CWE types do not have any recorded
real-world occurrences is due to their insignificance. Their presence in
real-world software systems may decrease the overall code quality, however they
do lead to detectable errors that would lead to a recorded vulnerability
instance. Another possibility is that errors due to such CWE types are more
likely to be detected by software testing tools, reducing the chance they end up
in the released product. For example, instances of most function related CWE
types, such as, CWE 685 - Function Call With Incorrect Number of Arguments,
would be easily detected by most software testing methods.

In \fig\ref{fig:overall_wordcloud}, we show the wordcloud of the description of mapped CVEs per the 12 CWEs for which we found at least one mapped CVEs. 
The themes depicted by the words in each wordcloud capture the weakness type and how attackers are exploiting those. For example, CWE-369 (Divide by zero) can be exploited by an 
attacker to stage a denial of service attack (as highlighted by words in the wordcloud like `denial', `cause', `zero', etc.). 
We now briefly describe the mapped CVEs along the six CWE weakness categories, i.e., Memory, Calculation, Conversion, Initialization, Function, and Resource. We discuss the 
categories in the order of the frequency of the mapped CVEs. In \tbl\ref{fig:highest-severity}, we provide an example of a mapped CVE per each of the 12 CWEs. Each example CVE is 
picked from the most severe CVEs mapped to the CWE (based on CVSS score). For example, for CWE-190 (Integer overflow), the CVE 2020-11279 has a CVSS score of 10 (i.e., most critical). 
The CVE is logged to capture incident like \emt{memory corruption due to improper length check in SDES packets}. We now briefly discuss the mapped CVEs based on the CWE weakness categories.

\nd\bf{2997 CVEs mapped to Memory type weaknesses - 5.6\%L 72.3\%M 16.1\%H 6.0\% C}\newline
The following 4 CWE types under the memory category have a total of 2997 CVE instances. Only 2 memory related CWE types did not have any CVE instances.

\textbf{CWE 190 - Integer Overflow or Wraparound} is mapped to \textbf{1418 CVEs from 5 Vulnerable Code Snippets}. As
shown in Table \ref{fig:highest-severity}, the most severe instance of this type
had a CVSS score of 10, indicating that it is a critical vulnerability. We
observe that this particular instance of integer overflow led to a significant
outcome of memory corruption. When we analyze the most
frequently occurring words in the descriptions of all 1418 instances, we see
that in Figure \ref{fig:overall_wordcloud} \subref{190wordcloud}, one keyword
that occurs frequently is "function". This indicates that integer overflow
vulnerabilities are often the result of poorly written functions, or improperly
calling memory management functions such as \it{malloc()}. We also observe that
the keyword "contract" occurs frequently, which refers to integer overflow
errors affecting smart contracts. \textbf{CWE 476 - NULL Pointer Dereference} is mapped to \textbf{1371 CVEs from 4 Vulnerable Code Snippets}.
The highest severity instance of CWE 476 - NULL Pointer Dereference (Table \ref{fig:highest-severity})
has a CVSS scores of 10 (i.e., critical). We observe that this instance of null-pointer dereference
occurred due to accessing a buffer beyond it's size, which ultimately lead to a
critical fault. In Figure \ref{fig:overall_wordcloud} \subref{476wordcloud}, we
see that they keywords "denial" and "service" are relatively common, indicating
that instances of null pointer dereference can lead to a denial of
service (DoS) attack. \textbf{CWE 401 - Failure to Release Memory} is mapped to \textbf{195 CVEs from 6 Vulnerable Code Snippets}. 
The highest severity instance has a score of 7.8 , which is considered to be high severity. 
As shown in Table \ref{fig:highest-severity}, this instance occurred to due to
memory leaks in functions, which lead to attacks potentially being permitted to
cause a denial of service (DOS), or excess memory consumption. We also observe
in Figure \ref{fig:overall_wordcloud} \subref{401wordcloud} that the keywords
"denial" and "service" occur frequently, indicating that DOS is a common
consequence of memory leaks. \textbf{CWE 788 - Access of Memory Location After End of Buffer} is mapped to \textbf{13 CVEs from 24 Vulnerable Code Snippets}.
This CWE type has the least number of CVE instances for Memory related vulnerabilities, 
yet has the highest occurrences in the Stack Exchange snippets. Although it may
not have as many real-world instances, we see that errors related to accessing
memory after end of a buffer can be easily introduced in forum answers.
In Table \ref{fig:highest-severity}, we observe that the highest CVE
instance is critical with a CVSS score of 9.3. This particular instance led a
potential memory corruption vulnerability if a user opened a malicious file (keywords also found in \fig\subref{788wordcloud}).

\nd\bf{239 CVEs mapped to Calculation type weaknesses - 14.2\%L 74.5\%M 9.2\%H 2.1\% C}\newline
The following 2 CWE types under the calculation category have a total of 239 CVE
instances. A total of 3 vulnerable code snippets contained these CWE types.

The \textbf{CWE 369 - Divide by Zero} is mapped to \textbf{196 CVEs in 1 Vulnerable Code Snippet}.
This CWE type has a relatively high number of CVE instances, but 
only 1 instance of CWE 369 was
observed in the Stack Exchange code snippets. The CVE instance 
with the highest severity had a
high severity CVSS scores of 7.8 (\tbl\ref{fig:highest-severity}). This particular vulnerable
instance resulted in attackers having the potential to cause a DOS via a divide
by zero error. From Figure \ref{fig:overall_wordcloud}
\subref{369wordcloud}, this is a common across 196 CVE (keywords "attacker", "denial", and "service").
The \textbf{CWE 682 - Incorrect Calculation} is mapped to \textbf{43 CVEs from 2 Vulnerable Code Snippets}.
As shown in Table \ref{fig:highest-severity}, it's highest instance is
considered to be critically severe with a CVSS score of 9.3. This instance resulted in a remote code execution
vulnerability. In Figure \ref{fig:overall_wordcloud} \subref{682wordcloud}
we observe that remote attacks are a common result as "remote" is a frequently
occurring keyword, along with denial of service vulnerabilities and elevation of
privileges. 

\nd\bf{165 CVEs mapped to Convesion type weaknesses  - 1.2\%L 78.8\%M 6.1\%H 13.9\% C}\newline
Only one out of the two CWE types that were related to conversion errors were able to be mapped to CVE instances. The 
\textbf{CWE 704 - Incorrect Type Conversion or Cast} is mapped to \textbf{165 CVEs from 1 Vulnerable Code Snippet}.
Around 13\% Vulnerabilities caused due to exploitation of this weakness are of critical nature.
The highest severity CVE instance has a CVSS scores of 10.  As shown in
Table \ref{fig:highest-severity}, this particular instance of a "type confusion"
error lead to arbitrary code execution, also corroborated by keywords like attacker, remote in \ref{fig:overall_wordcloud} \subref{704wordcloud}. 

\nd\bf{148 CVEs mapped to Initialization type weaknesses - 35.8\%L 46.6\%M 10.2\%H 7.4\% C}\newline
The following 2 Initialization CWE types have a total of 149 CVE instances. The \textbf{CWE 665 - Improper Initialization} is mapped to \textbf{148 CVEs from 3 Vulnerable Code Snippets}. 
It has 148 total CVE instances, which occur mainly due to improper memory initialization (see \fig\ref{fig:overall_wordcloud} \subref{665wordcloud}). 
The  most critical CVE has CVSS scores of 10. We see in Table \ref{fig:highest-severity}, this particular
improper initialization was related to improper memory initialization.
The \textbf{CWE 457 -  Use of Uninitialized Variable} is mapped to \textbf{1 CVE from 42 Vulnerable Code Snippets}. 
This CWE type has only one CVE instance, although it was the most frequently
detected among all Stack Exchange code snippets (42 compared to 3 for CWE 665 -
Improper Initialization). As shown in Table
\ref{fig:highest-severity}, the single CVE instance has a medium severity with a
CVSS score of 6.8, and lead to the software being at risk for arbitrary code
execution by attackers. 

\nd\bf{31 CVEs mapped to Function type weaknesses - 6.5\%L 61.3\%M 25.8\%H 6.4\% C}\newline
Although only there are a total of 7 different CWE types that we categorized as
Function related weaknesses, only one was able to be mapped to CVE instances.
\textbf{CWE 252 - Unchecked Return Value} is mapped to \textbf{31 CVEs from 3 Vulnerable Code Snippets}. 
It has 31 CVE instances with it's highest severity instance having a critically
severe score of 9.3. As shown in Table
\ref{fig:highest-severity}, this particular instance led to a code execution
vulnerability. Figure \ref{fig:overall_wordcloud}
\subref{252wordcloud} shows that code executions are not a common result, and there are
variety of consequences from unchecked return values. 

\nd\bf{14 CVEs mapped to Resource type weaknesses - 7.1\%L 35.7\%M 42.9\%H 14.3\% C}\newline
Both CWE types that we categorized under the Resource category have CVE
instances, with a total of 14. \textbf{CWE 672 - Operation on a Resource after Expiration or Release} is mapped to 
\textbf{12 CVEs from 1 Vulnerable Code Snippet}. 
This CWE has both a low number of CVE instances, and a low number of occurrences
in the Stack Exchange snippets. Table \ref{fig:highest-severity} shows that the
highest severity instance is of critical severity with a CVSS score of 9.3. We
observe that the outcome of this particular instance (privilege elevation) is a
common result of using an expired resource as Figure \ref{fig:overall_wordcloud}
\subref{672wordcloud} "privilege" is a frequently occurring keyword. \textbf{CWE 664 -  Improper Control of a Resource Through its Lifetime} 
is mapped to \textbf{2 CVEs from 2 Vulnerable Code Snippets}. Through its Lifetime has only 2 CVE instances
with the highest being of critical severity, as shown in Table
\ref{fig:highest-severity}. We see the outcome of improperly controlling the
release of a resource in this instance resulted in the potential for attacks to
cause a system restart or privilege elevation.
 
\begin{tcolorbox}[flushleft upper, boxrule=1pt, arc=0pt, left=0pt, right=0pt, top=0pt, bottom=0pt, colback=white, after=\ignorespacesafterend\par\noindent]
\noindent\textbf{Summary of RQ2: Mapping of CWE Types to CVE Instances} 
We find a total of 3595 CVE instances from 12 CWE types. These 12 CWE types belong to 6 
general weakness categories, with the majority belonging to the Memory category (4). 
CWE 190 -Integer Overflow or Wraparound (Memory category) contains the greatest number of CVE Instances with (1418).
\end{tcolorbox}

\subsection{RQ 3 How are the mapped CVE types classified/categorized in the CVE details database?}\label{sec:rq3}
\subsubsection{Motivation} While analyzing the CVE instances per the mapped CWEs in RQ2, we observed that 
the CWEs can cause diverse vulnerabilities like denial of service, resource lock, and so on. 
In total 3935 CVE instances are mapped to 12 CWEs that we observed in the shared IoT code examples. 
While as part of RQ2, we analyzed the 3995 CVEs along the eight code weakness categories from RQ1, 
it can also offer help if we analyze the mapped CVEs based on the type of vulnerabilities they can introduce to the system. 

As such, a high-level categorization of the 3995 CVEs would help us understand which categories are more prevalent 
in the shared code examples. 
\subsubsection{Approach} The cvedetails.com database contains detailed
information about each CVE instance, including it's categorization into one or
more of 13 different CVE types. These CVE types include general software
security vulnerabilities, methods of cyber attacks, and web application
exploitations.  
We first download the data set of CVE instances from
cvedetails.com for each analyzed CWE type from RQ1. Each entry in this data set
contains a CVE instance and its respective information, including its
categorization into one or more of the 13 CVE types used by cvedetails.com. We
then check the distribution of weak code snippets in our studied dataset along the CVE types. 

\subsubsection{Results} 
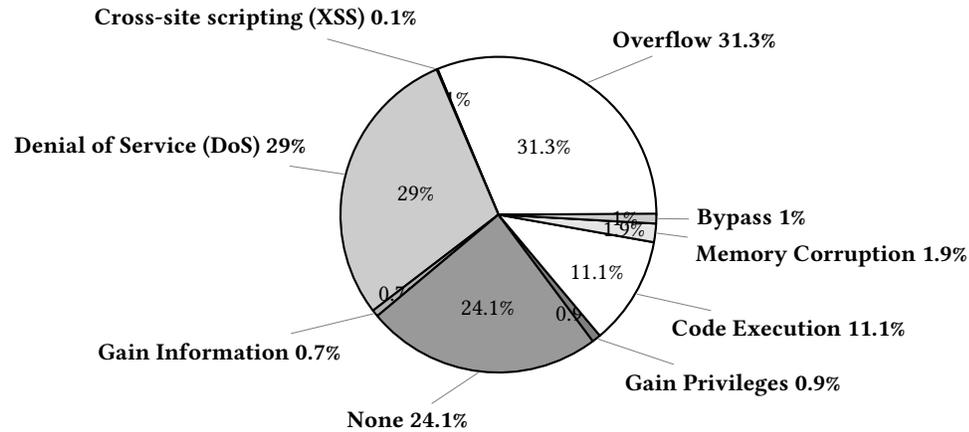
\begin{figure}[t]
    \centering\begin{tikzpicture}[scale=0.7]-
    \pie[
        /tikz/every pin/.style={align=center},
        text=pin,
        explode=0.0,
        color={black!0, black!10, black!20, black!30, black!40, black!50}
        ]
        {
            31.3/\bf{Overflow 31.3\%},
            0.1/\bf{Cross-site scripting (XSS) 0.1\%},
            29/\bf{Denial of Service (DoS) 29\%},
            0.7/\bf{Gain Information 0.7\%},
            24.1/\bf{None 24.1\%},
            0.9/\bf{Gain Privileges 0.9\%},
            11.1/\bf{Code Execution 11.1\%},
            1.9/\bf{Memory Corruption 1.9\%},
            1/\bf{Bypass  1\%}
        }
    \end{tikzpicture}
	\caption{Distribution of the mapped CVE instances to the 12 CWE types by the CVE Types}
	\vspace{-5mm}
	\label{fig:cve-cat}
\end{figure}
We observe that out of the 13 CVE types listed on
cvedetails.com, 
75.9\% of our 3559 mapped CVE instances belong to eight CVE types 
in the cvedetails.com database: \begin{inparaenum}[(1)]
\item Overflow, 
\item Denial of Service,
\item Code Execution,
\item Memory Corruption,
\item Bypass,
\item Gain Privilege,
\item Gain Information, and
\item Cross-site scripting.
\end{inparaenum} Among the 75.9\% CVEs, most belong to the Overflow type vulnerabiliites (31.3\%), followed by Denial of Service (29\%), and Code Execution (11.1\%).
Around 24.1\% of the mapped CVE instances are not assigned any CVE type in the codedetails.com database. \fig\ref{fig:cve-cat} shows the distribution of the mapped 
3995 CVE instances by their CVE types. The 5 excluded CVE types from cvedetails.com database in our IoT code examples are: Directory Traversal, SQL Injection, File
Inclusion, Cross-Site Request Forgery (CSRF), and Http Response Splitting.
Overall 12.45\% of the CVE instances in cvedetails.com's database belong to
these 5 types. Individually, each of these excluded CVE types do not contain a
significant amount of CVE instances. For example the Http Response Splitting
type only contains 0.1\% of all CVE instances in the database, and only 4 out of
171,015 instances recorded in 2021. As these CVE types and vulnerabilities are
largely concerned with Http security exploits and web applications, they are not
as likely to be present in C/C++ code.

The 8 CVE types that contain the detected CVE instances are broken down below:

\nd\textbf{Overflow Type Vulnerabilities: 7 CWE Types - 31.3\% of CVE Instances}\newline 
These CVE instances primarily involve improper calculations that lead to integer
overflow errors. Such errors can have severe consequences if the calculations
are security critical. We find that 7 out of the 12 CWE types with CVE instances
have instances that are of the type "Overflow".

\nd\textbf{Denial of Service (DoS) Type Vulnerabilities: 8 CWE Types - 29.0\% of CVE Instances}\newline
DoS vulnerabilities occur when a resource becomes unavailable. Eight of out the 12
CWE types contain DoS CVE instances. The CWE type with the greatest number is
CWE 476 - NULL Pointer Dereference (57.29\% or 691 out of 1213 instances). We
also find that out of the 4 CWE types with a significant number of DoS CVE
instances (greater than 100), 3 belong to the Memory type CWE category. These
CWE types are 476 - NULL Pointer Dereference, 190-Integer Overflow or
Wraparound, and Failure to Release Memory. Therefore we find that the majority
of Denial of Service CVE instances are due to memory related errors (86.72\% or
1052 out of 1213 instances). In a study on the security of smart cars, denial of
service was found to be a potential threat that could lead to unpredictable
behaviour in the car, and also communication failures ~\cite{smartcars}.

\nd\textbf{Code Execution Type Vulnerabilities: 9 CWE Types - 11.1\% of CVE Instances}\newline
Code execution vulnerabilities occur when a program unintentionally allows code
to be injected and executed. This can be a critical security flaw, as it allows
attackers to easily manipulate a program. We observe that this type of CVE
instance occurs in 9 out of the 12 CWE types, but primarily in CWE 190 -Integer
Overflow or Wraparound (48.49\%), CWE 704-Incorrect Type Conversion or Cast
(26.72\%), and CWE 476-NULL Pointer Dereference (15.51\%). CWE-704 belongs to
the Conversion CWE category, while CWE 190 and 476 are Memory related CWE types.
However, there are also a significant number of Initialization type
vulnerabilities with code execution type CVE instances (CWE 665 - Improper
Initialization with 22 out of 464 instances or 4.74\%). 

\nd\textbf{Memory Corruption Type Vulnerabilities: 4 CWE Types - 1.9\% of CVE Instances}\newline
These CVE instances occur when memory is modified due to unintentional behaviour
from the programmer. Some common causes include improper heap management or
using more memory than allocated. Such errors are common in code written in C
and C++ due to the need of explicit memory management. We observe that 4 out of
the 12 CWE types have memory corruption type CVE instances, with CWE 190
-Integer Overflow or Wraparound having the vast majority of instances (85\% or
68 out of 80 instances). Three out of the four CWEs belong to the "Memory" CWE
category. 

\nd\textbf{Bypass Type Vulnerabilities: 7 CWE types - 1\% of CVE Instances}\newline
Bypass vulnerabilities occur when authentication controls fail and attackers are
allowed to bypass and perform malicious operations. We find that 7 out of the 12
CWEs have bypass related CVE instances, with the following CWE types containing
the greatest number: CWE 190 -Integer Overflow or Wraparound (46.34\%), CWE
704-Incorrect Type Conversion or Cast (14.63\%), CWE 665 - Improper
Initialization (14.63\%). These CWE types belong to the Memory, Conversion, and
Initialization CWE categories respectively, suggesting that bypass related
vulnerabilities can occur from different types of software errors. 

\nd\textbf{Gain Privileges Type Vulnerabilities: 6 CWE Types - 0.9
\% of CVE Instances}\newline
These CVE instances occur when attackers are permitted to gain privileges to a
program, such as higher levels of access. We find 6 out of the 12 CWE have
vulnerabilities that allow for privileges and access to be exploited. These
instances primarily occur in CWE 190 -Integer Overflow or Wraparound and CWE
704-Incorrect Type Conversion or Cast (41.66\% and 36.11\% respectively). Both
belong to the Memory CWE category.

\nd\textbf{Gain Information Type Vulnerabilities: 6 CWE Types - 0.7\% of CVE Instances}\newline
This vulnerability primarily occurs when information is not properly protected
and is leaked to attackers. We this type of CVE instance in 6 out of the 12 CWE
types, with CWE 190 and CWE 476 containing the greatest number of instances
(40\% and 20\% respectively). As these two CWE types belong to the Memory CWE
category, we observe that memory related errors can contribute to important
program information being leaked.

\nd\textbf{Cross-site scripting (XSS) Type Vulnerabilities: 2 CWE Types - 0.1\% of CVE Instances}\newline
These attacks involve client side
scripts being injected into web pages. This type of vulnerability primarily
occurs in web applications. This CVE type occurred in only 0.07\% of all CVE
instance. We find that 2 out of the 12 CWE types (CWE 190 and 476) have XSS type
CVE instances, both with very few (2 and 1 respectively). 

\begin{table}[t]
  \centering
  \caption{Distribution  of CVE types by Stack Exchange Site.}
  \label{fig:cve_type_distribution}
    \begin{tabular}{p{4cm}p{3cm}p{3cm}p{1cm}p{1cm}p{1cm}}
    \toprule{}
    \textbf{CVE Type} & \textbf{\# Mapped CVEs} & \textbf{\# Mapped CWEs} & \multicolumn{3}{l}{\bf{\#CWEs per Stack Exchange Site}}\\
 & & &\#SO    & \#ARD &  \#RP \\
    \midrule
    Overflow	& 1308 & 7& 5&6&2\\
    Denial of Service (DoS)	& 1213 & 8 &  6&6&2		\\
    Code Execution	& 464 & 9 &  7&7&3		\\
    Memory Corruption	& 80 & 4 & 4&3&2		\\
    Bypass 	& 41 & 7 & 6&5&2 		\\
    Gain Privileges	& 36 & 6 & 5&4&2 		\\
    Gain Information	& 30 & 6 & 5&4&2 		\\
    Cross-site scripting (XSS)	& 3 & 2 & 2&2&2	\\
    \bottomrule
    \end{tabular}%
 
\end{table}
\begin{figure}[t]
    \centering
    \subfloat[\centering Top CWEs by \# of CVEs]{{\includegraphics[width=7cm]{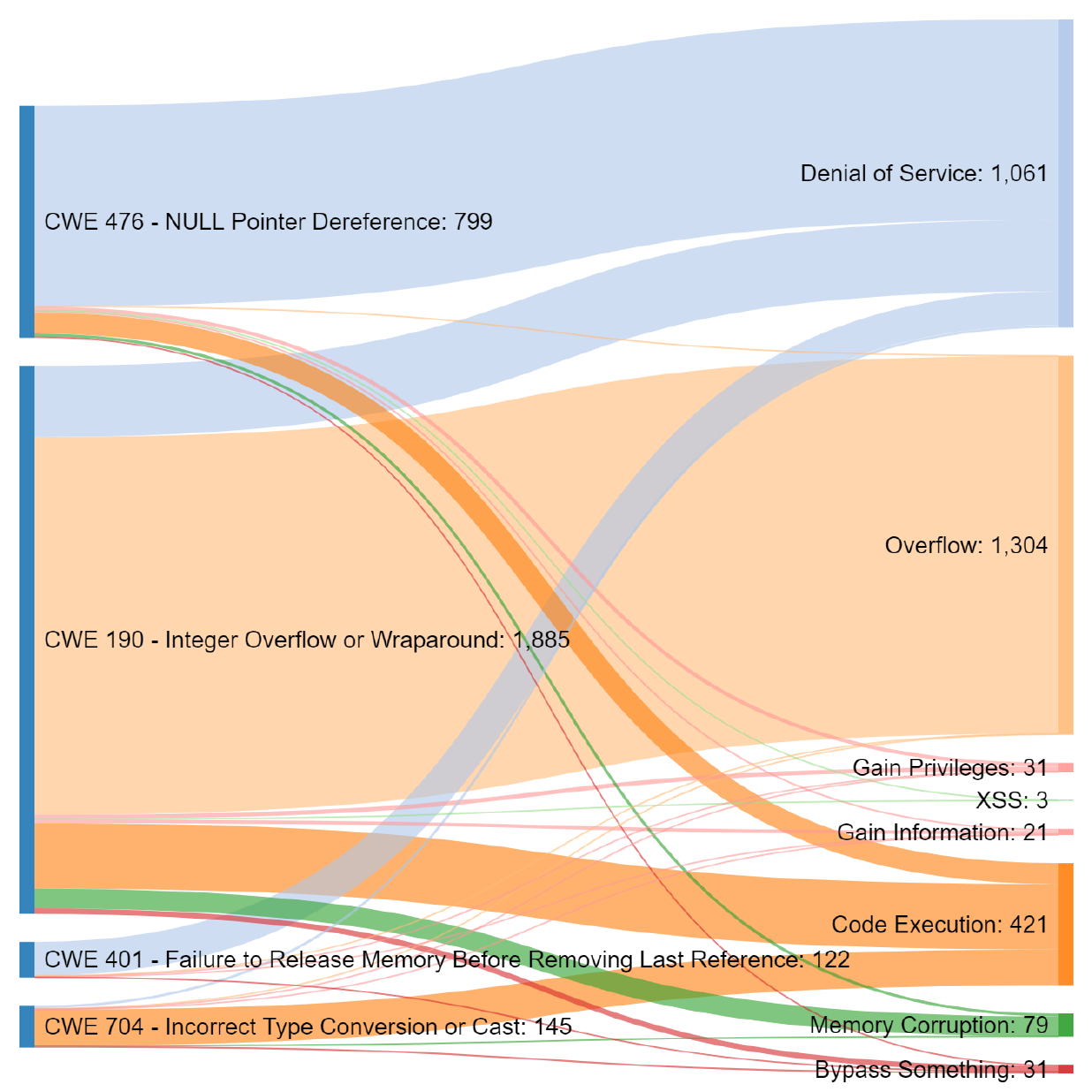} }}%
    \qquad
    \subfloat[\centering Bottom CWEs by \# of CVEs]{{\includegraphics[width=7cm]{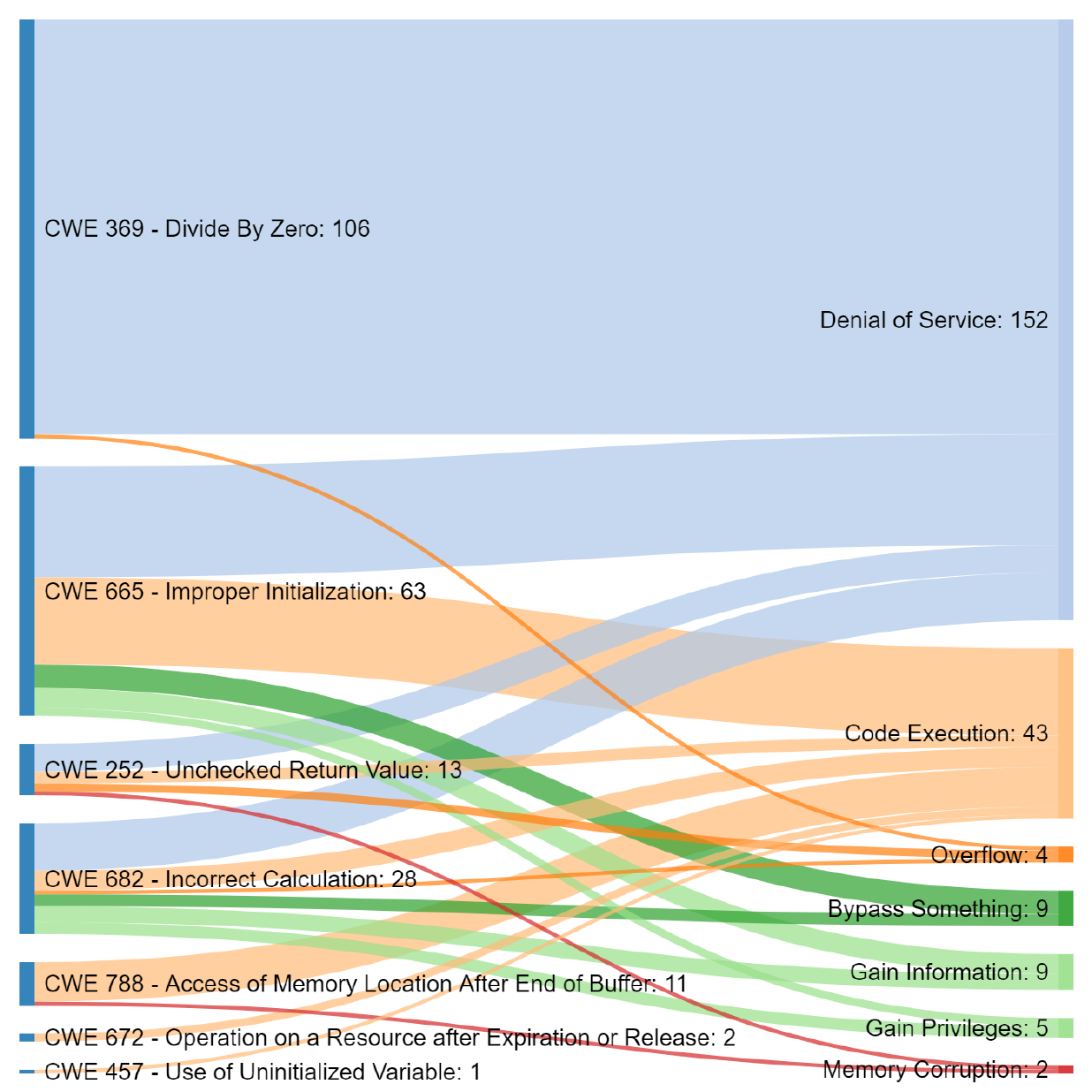} }}%
    \caption{Mapping of the found CWEs in the IoT Code to CVE Types. The width of each swim line denotes the total number of CVEs mapped to a CWE (i.e., larger width denotes more mapped CVEs)}%
    \label{fig:sankey-charts}%
\end{figure}
Overall, the 12 CWEs for which we mapped the CVE types, are distributed across all the three Stack Exchange sites, i.e., SO, Arduino, and Raspberry Pi. 
In \tbl\ref{fig:cve_type_distribution}, we show how the CWEs are distributed by grouping those by the CVE types. Out of our mapped 3995 CVEs, 
1308 belong to The CVE type `Overflow'. The 1308 CVEs correspond to 7 CWEs, where five are found in the code examples from SO, six from Arduino, and 
two from Raspberry Pi. This means that a CWE is observed in multiple sites and so the code examples in all the three sites can have the `Overflow' type 
vulnerabilities. Additionally, we observe in Table ~\ref{fig:cve_type_distribution} that there is not a significant difference in the distribution of the 
8 analyzed CVE types among the Stack Overflow, Arduino, and Raspberry Pi code
snippets. For the majority of the CVE types (6 out of 8), it's associated CWE
types occur more frequently in SO code snippets. For CVE types "Code
Execution" and "Memory Corruption", their associated CWE types occurred more
often in Arduino code examples.

In \fig\ref{fig:sankey-charts}, we show how a CWE can have vulnerabilities of multiple types. 
For example, if a code example shows integer overflow vulnerability (CWE 190), the weakness 
can be exploited to induce vulnerabilities related to `Denial of service', `Overflow', `Code Execution', and `Memory Corruption'. 

\begin{tcolorbox}[flushleft upper, boxrule=1pt, arc=0pt, left=0pt, right=0pt, top=0pt, bottom=0pt, colback=white, after=\ignorespacesafterend\par\noindent]
\noindent\textbf{Summary of RQ3: How are the mapped CVE types classified/categorized in the CVE details database?} 
The CVE instances we mapped to the detected CWE types in the shared Iot code
snippets belong to a total of 8 CVE types out of the 13 listed on
cvedetails.com. We observe that the most common CVE type is Overflow with
31.3\% of the CVE instances. A weakness in a code corresponding to a CWE can be exploited into different types of vulnerabilites. For example, 
the second most frequent weakness (CWE 190 - Integer Overflow) can lead four types of security vulnerabilites: Denial of service, Overflow, Code Execution, and Memory Corruption.
\end{tcolorbox}

\subsection{RQ 4 How do the IoT weaknesseses in the shared code examples evolve over time?}\label{sec:rq4}
\subsubsection{Motivation} The number of IoT related vulnerabilities that are
introduced in online Q/A forums varies each year as technology related to IoT is
constantly changing and being innovated. An analysis of the evolution of these
weaknesses in the studied forums can help us gain a better understanding as to which ones in particular
are occurring more frequently and should be paid more attention to.

\subsubsection{Approach} 
To determine the evolution trend of the identified CWE types, we first obtain and analyze the creation date of each code snippet that contained weaknesses. Using this information, we analyze the absolute and relative impact of the weak code snippets. We use the absolute impact refers to determine how the number of code snippets with weaknesses has been changing over time. In particular, we use the number of collected IoT snippets to determine a weakness category's absolute impact in a particular year. We also analyze the total absolute impact of all weak code snippets, as well as weak code snippets that don't contain instances of CWE 398 - Code Quality (i.e. snippets that only contain instances of the 28 distinct CWE types determined in RQ1). The absolute impact for a weakness category $C$ in a particular year $y$ is determined using the following formula: 
\begin{equation}
impact_{absolute}(C,y)= \frac{\# \textrm{Weak Code snippets}(C,y) }{\#\textrm{IoT Posts}(y)}
\end{equation}
Then to determine the total absolute impact in a particular year: 
\begin{equation}
impact_{total-absolute}(y)= \frac{\# \textrm{Weak Code snippets}(y) }{\#\textrm{IoT Posts}(y)}
\end{equation}
We then analyze the relative impact of the weakness categories, which involves analyzing the number of new weak code snippets within a specific category in relation to the total number of weak code snippets introduced in a particular year. Therefore, we use the following formula to determine the relative impact of a weakness category $C$ in a particular year $y$: 
\begin{equation}
impact_{relative}(C,y)= \frac{\# \textrm{Weak Code snippets}(C,y) }{\#\textrm{Weak Code Snippets}(y)}
\end{equation}
Furthermore, we analyze the relative change of the number of code snippets containing the 
28 distinct CWE types (i.e., non CWE-398 weaknesses) 
over a period of 12 years divided into 4 year groups. The relative change was determined using the following formula: 
\begin{equation}
\textrm{Relative Change}= \frac{\# \textrm{M present year group } - \# \textrm{M previous year group }}{\# \textrm{M previous year group }}
\end{equation}
\noindent where $\#M$ is the average number of weak code snippets introduced within a particular year group.
\subsubsection{Results}
\begin{figure}[t]
\centering
\includegraphics[scale=0.9]{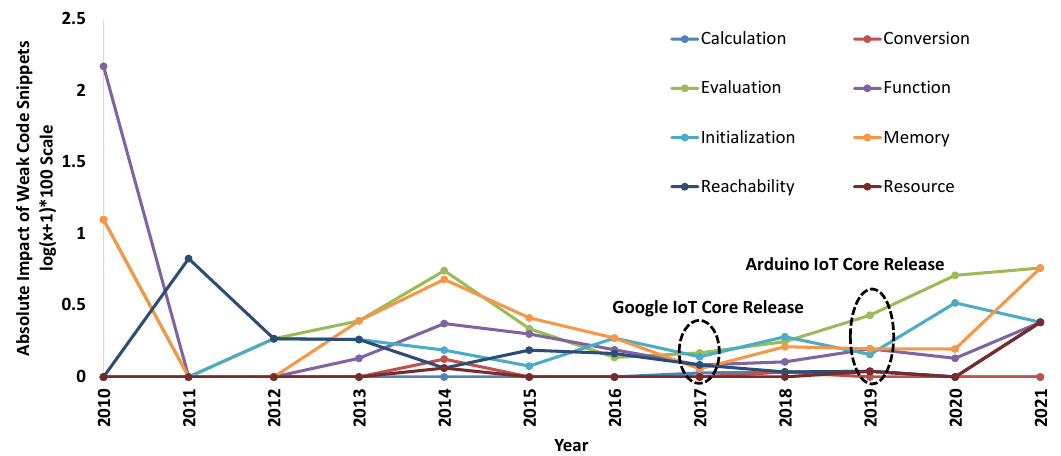}
\caption{Absolute Impact of Weakness Categories}
\label{fig:AbsoluteImpact}
\end{figure}

We first determine the absolute impact of the 8 weakness categories, and how it has been evolving over time. 
In Figure ~\ref{fig:AbsoluteImpact}, we observe that starting from 2014, there has been a downward trend in the number of weak code snippets across all weakness categories. Then starting from 2017, we observe a gradual increase. This is most apparent in code snippets related to evaluation, initialization, and memory errors. Then in 2019, we observe an increasing trend in the number of evaluation and initialization errors. Overall, we observe that in recent years, the number of weak code snippets is on the rise across most of the weakness categories. Correspondingly, in recent years new developments have been made to the field of IoT, which in turn has increased its popularity. Some examples of IoT related technology releases include Google Cloud's IoT Core management service in 2017 ~\cite{googleIoTCore}, and Arduino's release of their IoT Cloud application platform in 2019 ~\cite{arduinoIoTCloud}.

\begin{figure}[t]
    \centering
    \subfloat[\centering All CWEs]{{\includegraphics[width=7cm]{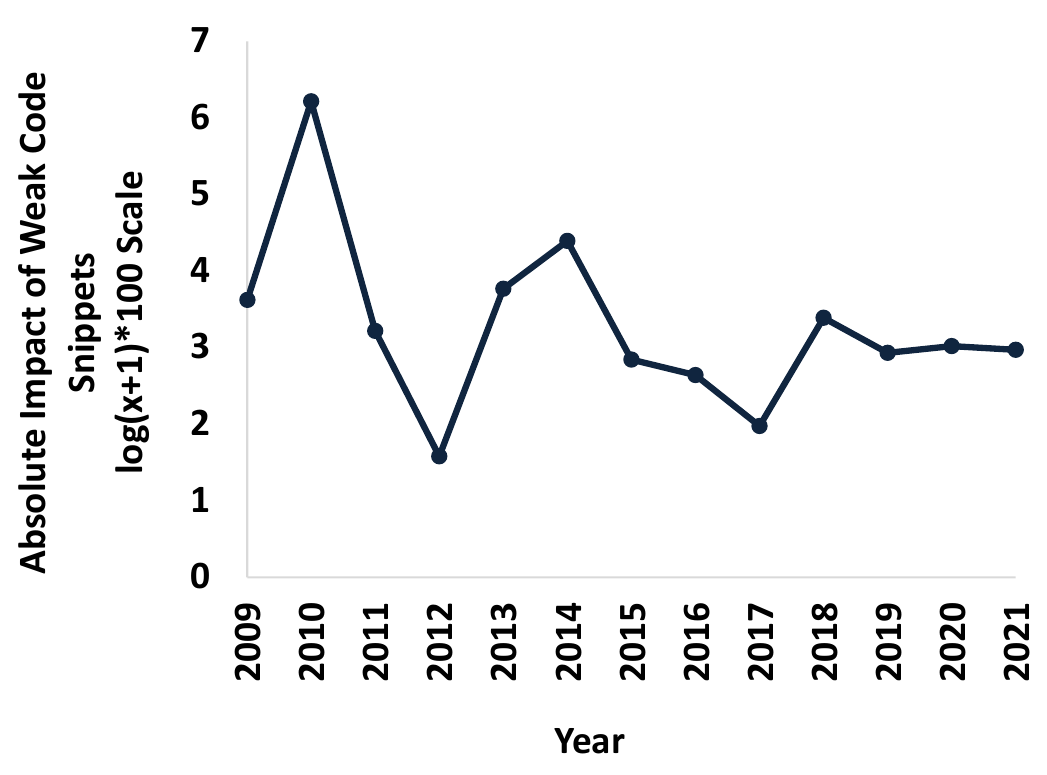} }}%
    \qquad
    \subfloat[\centering No CWE 398]{{\includegraphics[width=7cm]{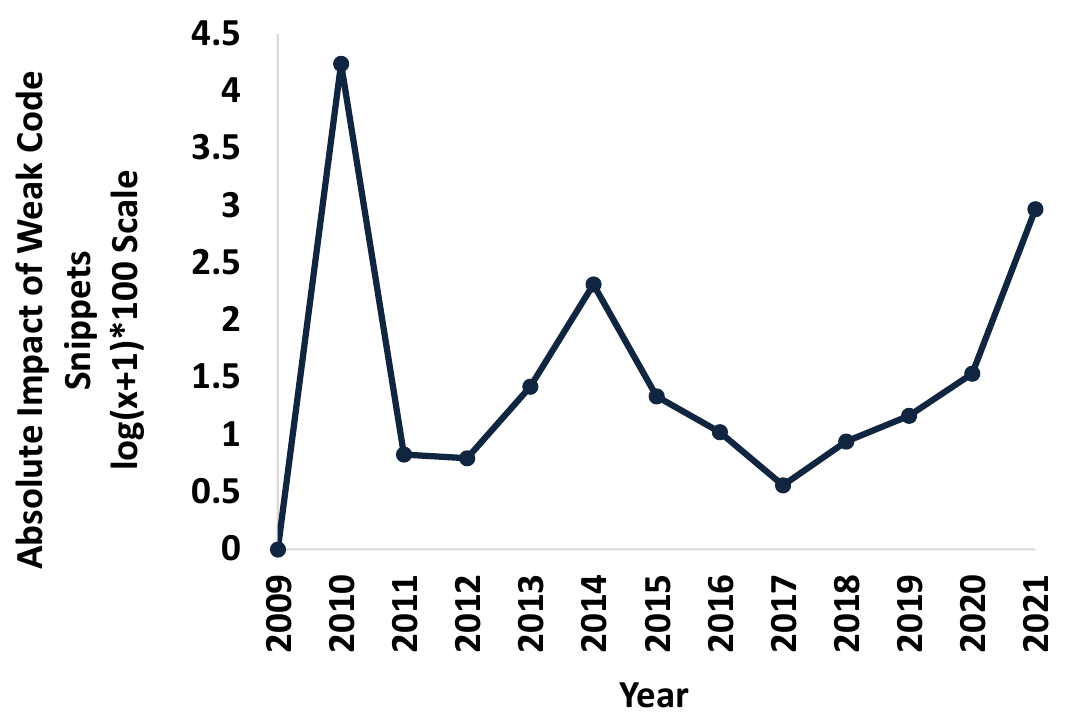} }}%
    \caption{Total Absolute Impact of Weak Code Snippets}%
    \label{fig:TotalAbsoluteImpact}%
\end{figure}

We then observe the differences in the total absolute impact of weak code snippets that contain instances of CWE 398 - Code Quality, and those that do not. Figure ~\ref{fig:TotalAbsoluteImpact} shows that the number of weak code snippets across all sites has been gradually decreasing since 2009. However, since 2018 the number of new code snippets with weakness that were introduced has not changed significantly. This indicates that the issue of weak code snippets on Stack Exchange sites is still persistent and requires attention. Furthermore, we observe that the number of code snippets that do not contain instances of CWE 398 has been steadily increasing since 2017. This is more concerning as these code snippets contain weaknesses that are of greater severity since CWE 398 is only an indicator of poor code quality. 

\begin{figure}[t]
\centering
\includegraphics[scale=0.9]{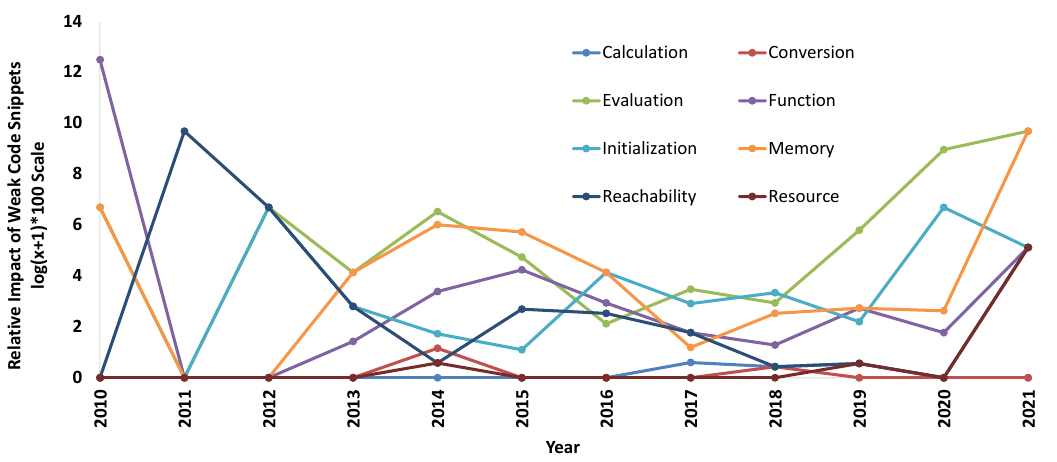}
\caption{Relative Impact of Weakness Categories}
\label{fig:RelativeImpact}
\end{figure}

In order to better understand the differences in the evolution trends among the 8 weakness categories, we also determine their relative impact to one another. As shown in Figure ~\ref{fig:RelativeImpact}, we find that relative to the other weakness categories, the number of code snippets containing initialization related errors decreased in 2020 while snippets with memory related errors increased. 

Finally, we analyze how the number of new weak code snippets has evolved within 4 year groups since 2009.  We observe in Table ~\ref{tab:relative_change} that the relative change of the average number of weak code snippets introduced within the 2018-2021 year group indicates a decline in weaknesses across the sites overall, and particularly in Stack Overflow and Raspberry Pi. However, we also observe that the number of new weak Arduino code snippets actually increased between 2018-2021, indicating that although the number of weak code snippets may be decreasing overall, it appears to remain a concern in the Arduino Stack Exchange site. 
\begin{table}[t]
    \centering
    \caption{Total and relative evolution of the weak code snippets (\#A = weak code snippets excluding those with only CWE-398, \#M = average \# of weak code snippets over the year group)}
\begin{tabular}{lrr|rrr|rrr|rrr}
\toprule
& \multicolumn{2}{c}{\bf{2009-2011}} &\multicolumn{3}{c}{\bf{2012-2014}} &\multicolumn{3}{c}{\bf{2015-2017}} &\multicolumn{3}{c}{\bf{2018-2021}}\\
\cmidrule{2-12}
 & \#A    & \#M &  \#A    & \#M &  $\delta $ & \#A    & \#M &  $\delta $ & \#A    & \#M &  $\delta $\\
\cmidrule{2-12}
All Sites & 5 & 1.7      & 52 &17.3 & 940  & 94& 31.3&  80.8  & 89 &22.3& -28.9\\
\midrule
Stack Overflow      & 5 & 1.7      & 30 &10 & 500 & 41 & 13.7&  36.7 & 21&5.3  & -61.6\\
Arduino     & 0 & 0      & 13  &4.3 &  100    & 44 & 14.7& 238.5 & 66&16.5  &  12.5\\
Raspberry Pi      & 0 & 0      & 9  &3 & 100   & 9 & 3&  0 & 2& 0.5  & -83.3\\
\bottomrule
\end{tabular}    
    \label{tab:relative_change}
\end{table}

\begin{tcolorbox}[flushleft upper, boxrule=1pt, arc=0pt, left=0pt, right=0pt, top=0pt, bottom=0pt, colback=white, after=\ignorespacesafterend\par\noindent]
\noindent\textbf{Summary of RQ4: How do the IoT weaknesses and vulnerabilities evolve over time?} 
Our findings show that although the absolute trend in the number of weak code snippets was in a decline between the years 2014-2017, it has been increasing in recent years, potentially due to innovations and developments to the field of IoT since 2017. In particular, we observe an increasing trend in the number of evaluation, initialization, and memory related errors. Across the studied Stack Exchange sites the number of new snippets with weakness has been decreasing. However, the number of weak Arduino code snippets has been steadily increasing in recent years. 
\end{tcolorbox}


\section{Implication of Findings}\label{sec:implications}
\begin{figure}[t]
    \centering
    \subfloat[\centering All CWEs]{{\includegraphics[width=7cm]{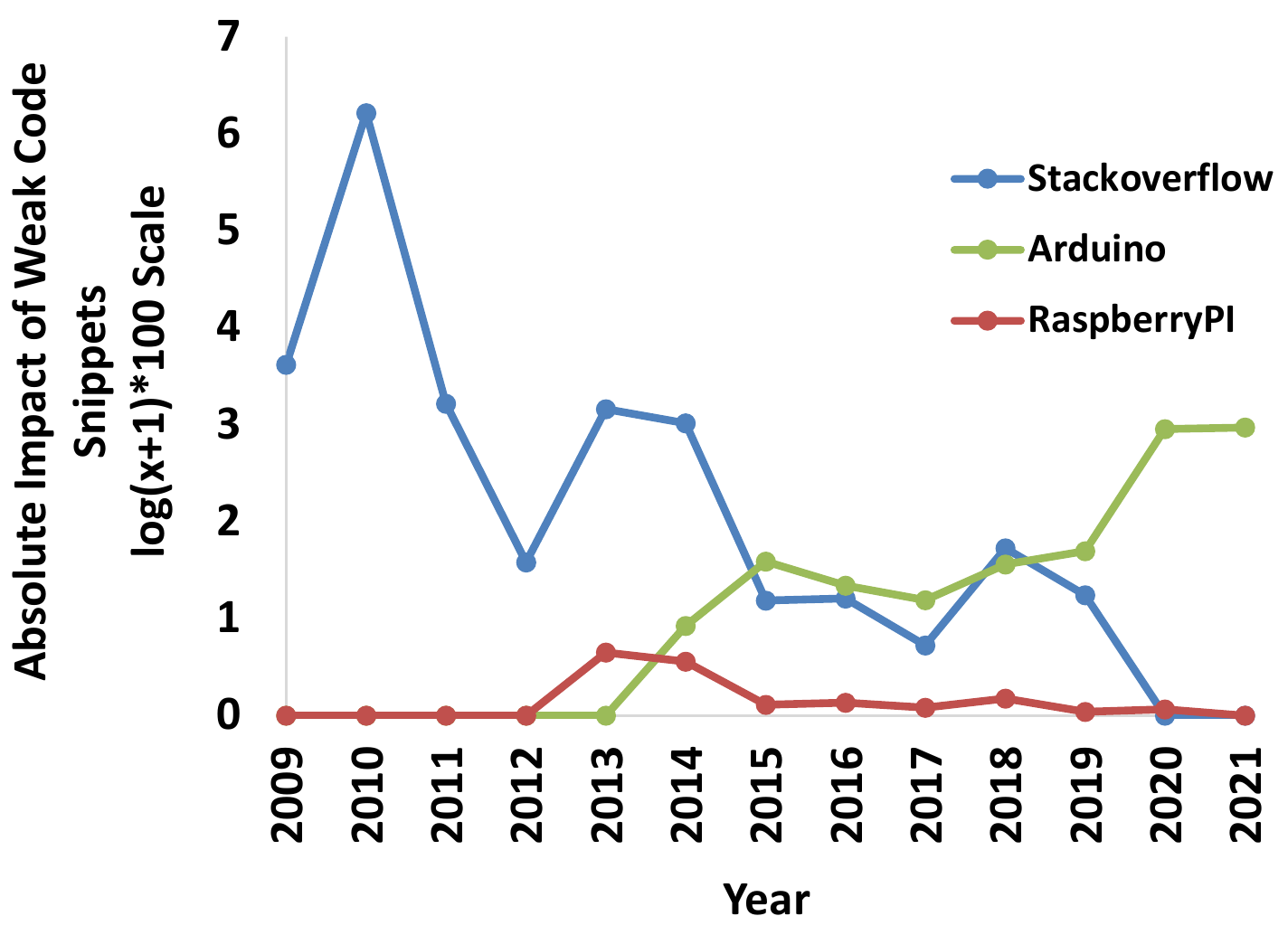} }}%
    \qquad
    \subfloat[\centering No CWE 398]{{\includegraphics[width=7cm]{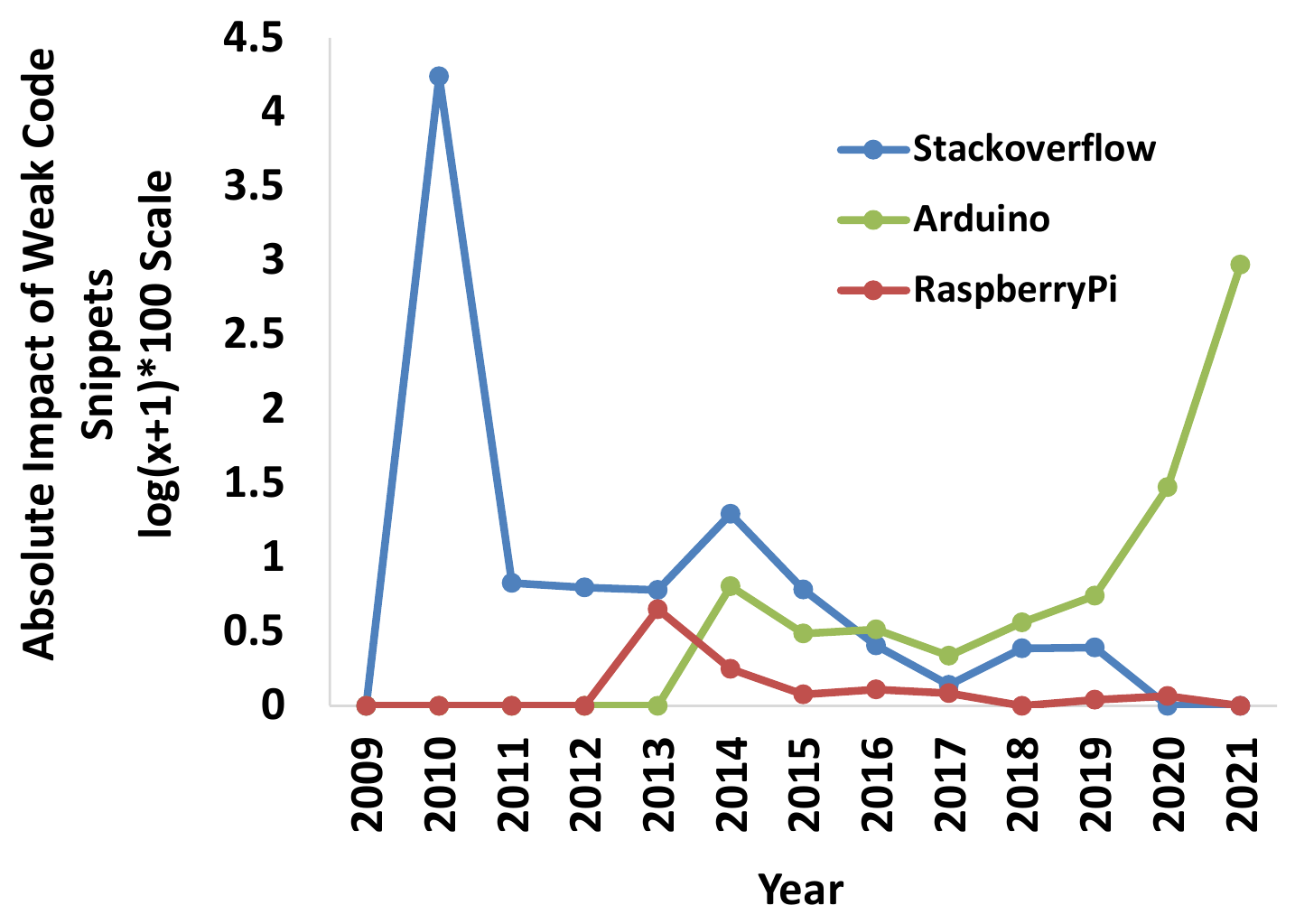} }}%
    \caption{Absolute Impact of Weak Code Snippets by Stack Exchange Site}%
    \label{fig:AbsoluteImpactBySite}%
\end{figure}
\begin{figure}[t]
\centering
\includegraphics[scale=0.80]{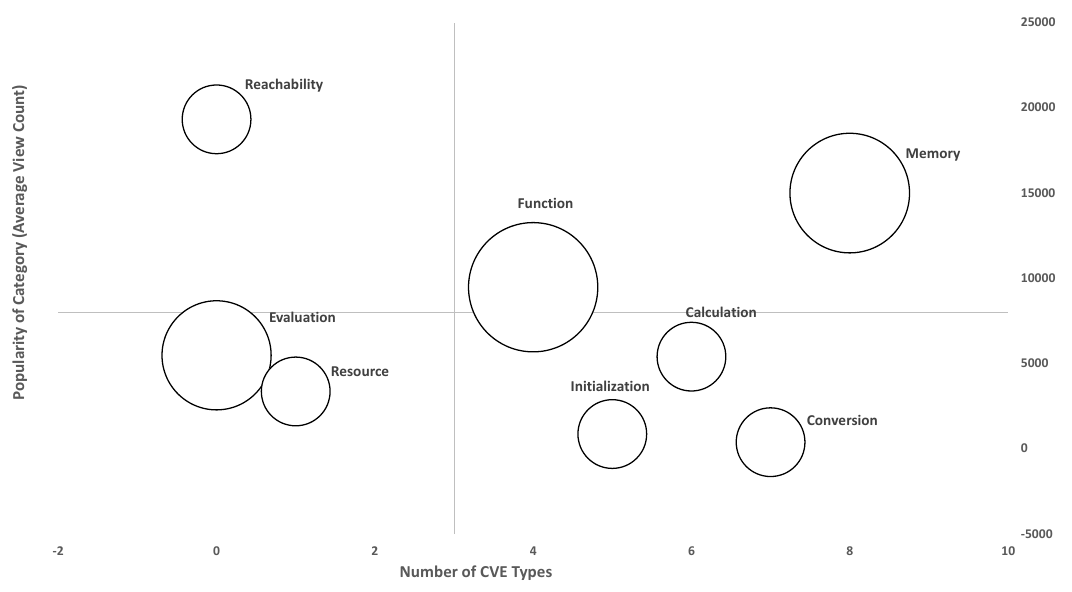}
\caption{Popularity of weakness categories compared to their real-world impact (number of CVE Types)}
\label{fig:bubble-chart}
\end{figure}

As previously mentioned in RQ4, the number of weak code snippets found on the
Arduino Stack Exchange site is increasing. As shown in Figure
~\ref{fig:AbsoluteImpactBySite}, this is true for  snippets that contain CWE 398
- Code Quality, and those that contain more severe types of weaknesses. When
compared to the evolution trends of Stack Overflow and Raspberry pi, there is a
significant different in the recent trends of Arduino. Therefore, the security
surrounding Arduino code snippets is of higher importance in regards to
understanding the risks associated with IoT development. 

The overall findings from this study are summarized in Figure
~\ref{fig:bubble-chart}, which shows the popularity and severity of the 8
general weakness categories based on the average view count and CVE types of
each category. The size of each bubble represents the to total number of CWE
types that are associated with a particular category.

We can use these findings and the findings from RQ1-RQ4 to guide the following IoT stakeholders: (1) IoT Developers, (2) IoT Educators, and (3) IoT Researches. 
We summarize the implications below.

\textbf{\underline{IoT Developers}:} IoT related products such as smart devices have
been increasing in popularity. As developers of these products often refer to
code examples posted on online Q/A while programming, they should be aware of
which common and harmful vulnerabilities exist in the examples. Figure
~\ref{fig:bubble-chart} shows that the Function, Memory, Evaluation categories
contain the largest number of CWE types. Weaknesses related to memory and
function errors in particular could be mapped to a large number of CVE
instances, and thus pose a greater security threat. Although there are no
evaluation related CWE types that could be mapped to any CVE instances, they are
still common in Stack Exchange code snippets and developers should be aware of
them while looking for programming solutions. Furthermore, the CVE instances
mapped to memory related CWE types could be grouped into 8 different CVE types.
Thus, improper memory management can lead to wide variety of real-world security
consequences, and developers must be aware of how memory is being handled in the
code examples they refer to, and make any necessary changes before they are
utilized. 

\textbf{\underline{IoT Educators}:} We see in Figure ~\ref{fig:bubble-chart} that
memory and function related weaknesses are high in number, and are associated
with a relatively large amount of CVE types (8 and 4 respectively). We also find
there is a large proportion of CVE instances mapped memory and function related
CWE types have either high or critical CVSS scores when compared to the other 6
types (22.1\% and 31.7\%). When educators are trying to teach or mentor others
on IoT topics or development, the focus should be on weaknesses that have the
highest potential security risk. Therefore, they should produce more educational
material on proper memory management and function use for developers in order to
increase their ability to both identify and resolve such vulnerabilities when
using online code examples.

\textbf{\underline{IoT Researchers}:} Certain code weaknesses have shown to be more
concerning due to their evolution patterns. Fig ~\ref{fig:AbsoluteImpact} in RQ4
shows that the trend of vulnerable code snippets being posted on Stack Exchange
sites is increasing for weaknesses related to evaluation, memory, and function
errors. Research is needed to further understand why some particular
vulnerability types continue to be introduced in these sites while others are no
longer as common. IoT researchers could ensure that vulnerabilities that are
increasing in frequency are identified on Stack Exchange sites to reduce the
chance of insecure code being copied. 

\section{Threats to Validity}\label{sec:threats}
\bf{Internal validity} threats relate to author bias in deciding which weaknesses to ignore, and how CWE types should be categorized. First, we notice in the initial cppcheck results that some claimed weaknesses were not accurate for this study. It is difficult to label weaknesses found in code segments as potentially harmful to users as there is no way of knowing how the segment will be used. For example, we do not know if a weakness will be automatically
addressed at another location in the user's project/software system. One way we
attempted to address this issue was by suppressing certain CWE types in
Cppcheck, such as CWE 563- Assignment to Variable without Use, however this
issue may still persist. Meaning that some of the vulnerabilities detected by
Cppcheck could be labeled as false positives. Another technique we followed from previous studies reduce bias is to suppress syntax errors identified by Cppcheck, and to ignore code snippets with less than 5 lines ~\cite{C/C++SO}. This decision ultimately lead to the removal of a large number of code snippets from our analysis. There is a possibility that these snippets could have contained further weaknesses that we did not analyze. However, suppressing syntax errors was necessary as there is a high change that they will be noticed by the user or their compiler, and we cannot claim such errors as harmful. Another potential threat to internal validity was the categorization of the 28 distinct CWE types into 8 weakness categories. This categorization was done by both authors and was revised multiple times after analyzing the root issues of the CWE types and their similarities to one anther. 

\bf{Construct validity} threats relate to errors that may have occurred during the collection of IoT related C and C++ code snippets from Stack Exchange sites. The data collection process varied by site. To determine if a Stack Overflow post was related to IoT, we looked for keywords present in the question tags. For Arduino and Raspberry, we made the assumption that all posts would be related to IoT. Another threat that concerns construct validity is our use of the language detection tool guesslang to identify C and C++ code snippets. The collection and analysis of non C/C++ code snippets in our study may have lead to inaccurate results. Guesslang has been used in previous studies to specifically detect C and C++ code, and has a validity rate of 90\% according to it's documentation ~\cite{guesslangdoc}. Furthermore, Zhang el al. found only a 10\% false positivity rate in their own manual testing of the tool ~\cite{C/C++SO}.   

\bf{External validity} threats relate to how our findings can be generalized to the nature of IoT posts on online Q/A sites as a whole. In this study we focused on vulnerabilities in Stack Exchange answers, not questions. This is because code snippets found in answers are meant to be "solutions", and are more likely to be copied and used by developers. However, code segments in the question itself could also contain vulnerabilities that may go unnoticed and replicated by the users attempting to provide answers to solve their problem. We also focus our study on strictly C and C++ code snippets due to their popularity in IoT development, although there may exist IoT related code examples in other languages that may have their own unique security concerns. 


\section{Related Work}
We first compare our study results with two most related work that also studied weaknesses in the C/C++ code examples shared in Stack Overflow (\sec\ref{sec:comparison-related-work}). We then summarize other related work in \sec\ref{sec:other-related-work}. 
\subsection{Comparison with Previous Studies Checking Weaknesses on Shared C/C++ Code Examples}\label{sec:comparison-related-work}
To the best of our understanding, the C/C++ code examples shared in Stack Overflow were subject to two empirical 
studies recently, first by Verdi et al.~\cite{EmpiricalC++Study} and then by Zhang et al.~\cite{C/C++SO}. 
Our study differs from the two studies as follows.
\begin{enumerate}
  \item Besides analyzing code examples from Stack Overflow, we also analyzed C/C++ code examples shared in four major Stack Exchange sites for IoT: Arduino, Raspberry Pi, IoT, and IoTa. 
  \item While the previous two studies focused on all types of C/C++ code examples, we focused on only the IoT related C/C++ code examples. 
  \item Unlike the previous two studies, our research questions are different as follows. 
  \begin{enumerate}
  \item We produce a category of the observed weakness types. A categorization is absent in the previous papers.
  \item We show how the weaknesses can be mapped to different CVE types (e.g., Overflow, Denial of Service, etc.).
  \item We analyze the evolution of the weaknesses based on their absolute and relative impacts.
  \item We conduct the analyses across all the three Stack Exchanges sites, where weaknesses are observed: Stack Overflow, Arduino, and Raspberry Pi,
  \end{enumerate}   
\end{enumerate} We summarize some of the similarities and differences between our results and previous studies below.

As previous studies have also analyzed vulnerabilities in C/C++ code snippets
found on online Q/A sites, there are some similarities and differences in the
results obtained. First, in a study of C/C+ code snippets found on Stack
Overflow by Zhang et al., 32 CWE types out of the 89 related to C/C++ were
detected ~\cite{C/C++SO}. Verdi et al. found 31 CWE types in their study of
C/C++ code examples from Stack Overflow ~\cite{EmpiricalC++Study}. Both of these
numbers are similar to our results of 28 distinct CWE types identified in code
snippets from Stack Overflow, Arduino, and Raspberry Pi. Furthermore, Zhang et
al., who similarly used cppcheck to automatically detect CWE instances, found
that 1.82\% of their collected code snippets (11,748 out of 646,716) contained
weaknesses. This is similar to our results of 2.10\%, or 240 out of the 11,329
obtained code snippets containing CWE instances. However, Verdi et al. did not
use a static code analyzer to detect vulnerabilities in their code snippets, and
instead manually reviewed 72,483 code snippets through multiple rounds. They
found vulnerabilities in 99, or 0.14\% of their Stack Overflow code snippets.

In terms of the types of vulnerabilities detected, Zhang et al. detected 16 of
the same CWE types as found in our study. Out of these 16 types, we observe that
CWE 788 - Access of Memory Location After End of Buffer was the most frequently
detected. The most common general weakness categories that these 16 types belong
to is the memory category with 5 CWE types, and the function category with 3 CWE
types. In total, 13 CWE types were detected in our analysis but not by Zhang et
al. Overall, these 13 types are categorized as mostly function (4 CWE types) and
evaluation (3 CWE types) related weaknesses. Although Zhang et al. did observe
CWE types that belong to these categories, such as CWE 131- Incorrect
Calculation of Buffer Size which can be categorized as an evaluation weakness,
they did not observe the exact weakness. One possible reason why Zhang et al did
not observe these CWE types is that they are less likely to occur in code
examples posted to stack overflow, which is the case for 4 out of the 13 CWE
types. In particular, two evaluation related weakness, CWE 570 - Expression is
Always False and CWE 595- Comparison of Object References Instead of Object
Contents, were more common in Arduino code snippets.
The pattern of CWE types detected by Zhang et al. that are different in their
exact definition, but can still categorized into one of our general weakness
categories is apparent when we observe their most frequently occurring CWE type,
CWE 908 - Use of an Uninitialized Resource (in 54.2\% of their snippets).
Although not an exact match, it is similar in nature to the most frequently
occurring CWE type in our analyzed snippets, CWE 457 - Use of Uninitialized
Variable (17.5\%), as they both involve improper initialization.

When we analyze the results obtained by Verdi et al., we notice that only 9 CWE
types that were found in our study were also observed in theirs. Out of these 9,
CWE 686 - Function Call With Incorrect Argument Type was the most frequently
occurring. Similar to our comparison with Zhang et al., these 9 CWE types are
the result of mostly function and memory related errors. From the 20 CWE types
we observed but were not detected by Verdi et al., we notice another similar
trend in that most of these CWE types are related to evaluation errors (5 in
total). These results are summarized in Table ~\ref{table:results_comparison}.
\begin{table}[ht]
        \centering
        \caption{Summary of results from previous studies compared with results of this study}
        \begin{tabular}{lrr}
        \toprule
         & \bf{Distinct CWE Types Detected} & \bf{Similar CWE types to this study}\\
        \midrule
        Zhang et al.~\cite{C/C++SO} & 32 & 16\\
        Verdi et al.~\cite{EmpiricalC++Study} & 31 & 9\\
        \bottomrule
        \end{tabular}
        
        \label{table:results_comparison}
\end{table}

Zhang et al. analyzed CVE instances of their detected CWE types to better
understand the potential impact of the vulnerabilities. They observe that 12 of
the 32 CWE types they detected could be mapped to CVE instances, similar to our
observations of 12 out of the 28 distinct CWE types having CVE instances.
Additionally, they found that 62.5\% of their detected CWE types did not have
any recorded CVE instances, while we found 57.14\% of our CWE types not have CVE
instances. This shows that in both IoT related code snippets obtained from three
Stack Exchange sites, and in Stack Overflow code examples that aren't related to
any one topic, the majority of vulnerabilities do not belong to CWE types that
have practical impact on real world software systems. 

Although Zhang et
al. studied the CVE instances that could be mapped to their detected CWE types,
they did not further analyze which CVE type the instances fell under. Analyzing
this information gave us more insight on the specific consequences of
vulnerabilities occurring in real world software. We observed that the 3595 CVE
instances we mapped from the detected CWE types were most frequently of the type
Overflow, Denial of Service, Code Execution. Therefore, these security flaws and
attacks occur the most often when instances of the detected vulnerabilities
exist in real-world software. Furthermore, unlike the previous studies where
observations were made each individual CWE type detected, in our analysis we
observe the impacts and characteristics of our detected CWE types within the
general weakness category they belong. By grouping the individual CWE types into
8 general categories, we obtained further information on the types of errors
users commonly introduce when posting programming solutions. Ultimately, we find
that users are more likely to make errors related to poorly written functions,
improper memory management, and incorrect evaluations.

\subsection{Other Related Work}\label{sec:other-related-work}
\begin{table}[ht]
  \centering
  \caption{Comparison between our study and previous related work}
    \begin{tabular}{p{1.5cm}p{4cm}|p{4cm}|p{4cm}}\toprule
    \textbf{Theme} & \textbf{Our Study} & \textbf{Prior Study} & \multicolumn{1}{l}{\textbf{Comparison}} \\ 
    \midrule
    \bf{Types of code snippets analyzed} & We look analyze code snippets from 3 Stack Exchange sites (Stack Overflow, Arduino, and Raspberry Pi). A total of 11,329 C/C++ code snippets were obtained from questions that pertain to IoT and embedded projects. & 
Previous research has analyzed C and C++ code snippets from Stack Overflow posts ~\cite{C/C++SO} ~\cite{EmpiricalC++Study}. The posts are not limited by topic and a wider variety of code examples are analyzed for vulnerabilities. A larger number of snippets are also analyzed in previous studies, with some analyzing 646,716 different posts ~\cite{C/C++SO}.
    & We analyze code snippets from a variety of sites that are related to IoT. We do not examine every post that contains C/C++ code and instead limit our study to those that pertain to IoT. 
 \\
    \midrule
    \bf{Analysis of the types of vulnerabilities present in code snippets} & We analyzed Stack Exchange code snippets for vulnerabilities using cppcheck. The CWE types identified by cppcheck were then further categorized into 8 general weakness categories based on common characteristics. Our study continued to analyze the vulnerable code snippets within the 8 general categories.  &Previous studies use static code analyzers such as cppcheck to detect instances of CWE types in Stack Overflow code snippets, or analyzed code examples for vulnerabilities manually. Characteristics of each identified CWE type, such evolution and number of revisions, were studied individually ~\cite{C/C++SO}. Other characteristics, such as prevalence in github projects, were also studied within each CWE type ~\cite{EmpiricalC++Study}. 
    & We analyze all vulnerabilities detected in the Stack Exchange code snippets within their respective weakness category. Instead of performing certain evaluations, such as examining their evolution trends, on each CWE type individually, we analyze them within their general weakness category.\\
     \midrule
    \bf{Mapping of CWE types to CVE instances}& Each identified CWE type that occurred in the code examples was mapped to their respective CVE instances. We further which CVE types these instances fall under. Then, we analyze the number of CVE instances within each category, as well as the distribution of CVSS scores and the evolution of the CVE types. & Previous research examined how the number of CVE instances correspond to a CWE type's potential impact ~\cite{C/C++SO}. Median CVSS scores of the instances were used to measure severity, and the trend of the number of yearly instances was also analyzed. 
    & We analyze the mapped CVE instances in detail by determining which types of instances, such as Denial of Service or Code Execution, occur more frequently when certain CWE types occur in real life software. Furthermore, we also look at common keywords in the descriptions of CVE instances for common errors made that led to a vulnerability.\\
\\
    \bottomrule
    \end{tabular}%
  \label{related-work-table}%
\end{table}%
Other related work is summarized in Table ~\ref{related-work-table} based on the following categories: 1) Types of code snippets analyzed , 2) Analysis of the types of vulnerabilities present in code snippets, and 3) Mapping of CWE types to CVE instances. It can also be broadly divided into \bf{Studies} and \bf{Techniques} to understand and mitigate IoT related security issues.

\nd\bf{\underline{Studies.}} Previous studies on IoT have studied underlying middleware solutions (e.g.,
Hub)~\cite{Chaqfeh-ChallengesMiddlewareIoT-2012}, the use of big data analytics
to make smarter devices~\cite{Marjani-IoTDataAnalytics-IEEEAccess2017}, and the
design of secure protocols and
techniques~\cite{Fuqaha-IoTSurveyTechnologiesApplications-IEEECST2015,Khan-IoTSecurityReview-FGCS2018,Zhang-IoTSecurityChallenge-SOCA2014}
and their applications on diverse domains (e.g.,
eHealth~\cite{Minoli-IoTSecurityForEHealth-CHASE2017}). We are aware of no previous papers that have conducted a qualitative analysis of vulnerable code examples obtained from a variety of online Q/A sites. Stack Overflow posts have been previously studied for insecure python vulnerabilities ~\cite{PythonVulnerabilities}, topics discussed by IoT developers ~\cite{Uddin-IoTTopic-EMSE2021}, the prevalence of machine learning being adopted into IoT ~\cite{uddin2021security}, big
data~\cite{Bagherzadeh2019} and chatbot issues~\cite{abdellatifchallenges}. Zhang et al. ~\cite{C/C++SO}, and Verdi et al. ~\cite{EmpiricalC++Study} have studied C/C++ code snippets on Stack Overflow, but did not expand their study to other Stack Exchange sites, and did not focus their analysis on any specific topic such as IoT.  

\nd\bf{\underline{Techniques.}} IoT devices 
can be easy targets for cyber threats~\cite{Zhang-IoTSecurityChallenge-SOCA2014,Frustaci-IoTSecurityEvaluation-IEEEIoTJournal2017}. As such, significant research efforts are underway to improve IoT security. Automated IoT security and safety measures are
studied in Soteria \cite{Celik-IoTSafetySecurityAnalysis-USENIX2018},
IoTGuard~\cite{Celik-IoTDynamicEnforcementOfSecurity-NDSS2019}. Encryption and
hashing technologies make communication more secure and certified~\cite{Tedeschi-LikeSecureIoTCommunications-IEEEIoT2020}. Many authorization
techniques for IoT are proposed like SmartAuth~\cite{YuanTian-APIBot-ASE2017}. For smart home security, IoT security techniques are proposed like Piano~\cite{Gong-IoTPIANO-ICDCS2017},  smart authentication~\cite{He-RethinkIoTAccessControl-USENIX2018}, and cross-App Interference threat mitigation~\cite{Chi-SmartHomeCrossAppInference-DSN2020}. Session management and token verification are used in web security to
prevent intruder getting information. Attacks on Zigbee, an IEEE specification used to support interoperability can make IoT devices vulnerable~\cite{Ronen-IoTNuclearZigbeeChainReaction-SP2017}. Our study provides insight on weaknesses present in IoT related Stack Exchange posts. IoT developers that visit these sites can benefit from being aware of which weaknesses are most common, and the security risks associated with them. 

\section{Conclusions} \label{sec:conclusion}
The rapid expansion of IoT applications, and the popularity of Stack Exchange sites for programming solutions raises concern over the nature of code weaknesses present in IoT related posts. In this study, we analyzed 11,329 code examples from the Stack Overflow, Arduino, and Raspberry Pi Stack Exchange sites. Overall we found a total of 29 CWE types present in 609 code snippets. We observed that weaknesses related to improper functions, evaluations, and memory management are the most common. Additionally, we observed that CWE types that are related to memory errors are relatively common in both the analyzed snippets, and in real-world software systems as they can be mapped to a large number of CVE instances. These mapped CVE instances can be linked to Denial of Service (DoS), overflow, and code execution vulnerabilities. When we analyzed the evolution of the 8 weakness categories, we found that code snippets containing weaknesses related to those related to evaluation, initialization, and memory error have been experiencing an increasing trend. The security of Arduino code snippets is of more concern as the number of new weak code snippets has been significantly increasing in recent years compared to Stack Overflow and Raspberry Pi. The results from our study can be used by diverse IoT stakeholders to stay aware of the IoT security concerns found in crowd-shared code examples and to prioritize the development of tools and techniques to mitigate the concerns. Our future work aims to understand the human factors associated to the sharing and usage of the insecure code like whether and how the activity/expertise of the IoT developers may be correlated to their sharing of insecure code examples.

\begin{small}
\bibliographystyle{abbrv}
\bibliography{consolidated}
\end{small}
\end{document}